\documentclass[12pt]{iopart}

\usepackage{graphicx}
\usepackage{dcolumn}
\usepackage{bm}

\begin{document}

\title{Trapped Rydberg Ions: From Spin Chains to Fast Quantum Gates}

\author{Markus M\"uller$^1$, Linmei Liang$^2$, Igor Lesanovsky$^1$ and Peter Zoller$^1$}
\address{$^1$Institute for
Theoretical Physics, University of Innsbruck, and Institute for
Quantum Optics and Quantum Information of the Austrian Academy of
Sciences, Innsbruck, Austria}
\address{$^2$Department of Physics, National University of Defense
Technology, Changsha 410073, China}
\ead{markus.mueller@uibk.ac.at}

\date{\today}

\begin{abstract}\label{txt:abstract}
We study the dynamics of Rydberg ions trapped in a linear Paul trap, and discuss the properties of ionic Rydberg states in the presence of the static and time-dependent electric fields constituting the trap. The interactions in a system of many ions are investigated and coupled equations of the internal electronic states and the external oscillator modes of a linear ion chain are derived. We show that strong dipole-dipole interactions among the ions can be achieved by microwave dressing fields. Using low-angular momentum states with large quantum defect the internal dynamics can be mapped onto an effective spin model of a pair of dressed Rydberg states that describes the dynamics of Rydberg excitations in the ion crystal. We demonstrate that excitation transfer through the ion chain can be achieved on a nanosecond timescale and discuss the implementation of a fast  two-qubit gate in the ion chain.
\end{abstract}






\maketitle
\section{Introduction}
Rydberg states correspond to the infinite series of excited bound
states in a Coulomb potential with large principal quantum number
$n$. In view of their ``macroscopic'' size, $a_{\mathrm{Ry}}\sim n^{2}a_{0}$
with $a_{0}$ the atomic Bohr radius, Rydberg states have remarkable
properties, as reflected, for example, in their response to external
static and time dependent electric and magnetic fields \cite{Gallagher}. While the single particle physics of Rydberg atoms has been the subject of intensive
studies in the context of laser spectroscopy, recent interest has
focused on exploiting the large and long-range interactions between
laser excited Rydberg atoms to manipulate the many-body properties
of cold atomic ensembles. Examples include recent seminal experiments
on frozen Rydberg gases obtained by laser excitation from cold atomic
gases, demonstrating in particular a dipole-blockade mechanism \cite{Lukin01,Amthor07,Heidemann07,Raitzsch08,Heidemann08,Tong04}, which in sufficiently dense gases prevents the excitation of ground state atoms in the vicinity of a Rydberg atom, and proposals for fast two-qubit quantum gates between pairs of atoms in optical lattices \cite{Jaksch00}. Furthermore, Rydberg atoms have been proposed to serve as model systems for studying coherent transport of excitations \cite{Mulken07} - a mechanism which is of great importance for coherent energy transfer in biological systems, e.g.~in light-harvesting complexes \cite{Olaya-Castro08,SenerSchulten}. While these investigations have so far concentrated on neutral atoms, we are interested below in describing the properties of laser excited Rydberg ions stored in a Paul trap, in particular the interplay between trapping fields and Rydberg excitations, and the associated many-body interactions in a chain of cold trapped Rydberg ions.

\begin{figure}\center
\includegraphics[angle=0,width=8cm]{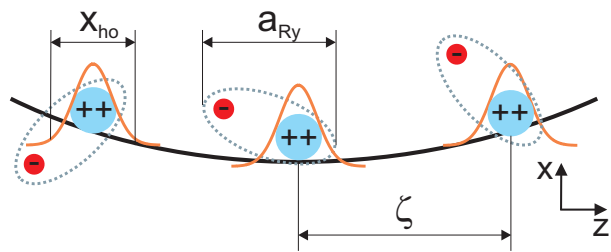}
\caption{Typical length scales in a chain of cold Rydberg ions in a linear Paul trap. The external trapping frequency is in the order of MHz with a corresponding oscillator length $x_\mathrm{ho}$ of approximately 10 nm. The interparticle spacing $\zeta$, set by the equilibrium between the Coulomb forces among the ions and the external confinement, is typically about 5 $\mu\mathrm{m}$. The third length scale is the size of the Rydberg orbit $a_\mathrm{Ry}$. Due to the scaling proportional to the square of the principal quantum number $n$ it can assume values in the order of 100 nm and therefore become significantly larger than $x_\mathrm{ho}$. In this regime the Rydberg ion cannot be considered as a point particle but rather as a composite object, and its internal structure must be taken into account.}\label{fig:lengthscales}
\end{figure}

Atomic ions can be stored in static and radio frequency (RF) electric quadrupole fields constituting a Paul trap \cite{Leibfried03}, where for sufficiently low temperature they form a Wigner crystal \cite{Bluemel88,Itano98,James98,Porras04}. Using laser cooling the ions can be prepared in the vibrational ground states of the phonon modes of the crystal. Internal electronic states of the ions can be manipulated with laser light and entangled via the collective phonon modes. Here we consider a situation where ions initially prepared in their electronic ground state are excited with laser light to high lying Rydberg states. In contrast to ions in low lying electronic states, a Rydberg ion in a Paul trap must be understood as a \emph{composite object}, where the Rydberg
electron is bound by the Coulomb force to the doubly charged ion core,
but both the Rydberg electron and the ion core move in the electric
fields constituting the trapping potentials. This is reminiscent to the situation in which Rydberg atoms are confined by very tight magnetic traps \cite{Lesanovsky05,Hezel06}. We are particularly interested in the parameter regime where the size of the Rydberg orbit $a_{\mathrm{Ry}}$
is larger than the localization length of the doubly charged ion core
$x_\mathrm{ho}$ around its equilibrium position, but still much smaller than
the average distance $\zeta$ between the ions in the Wigner crystal,
i.e.~$x_\mathrm{ho} < a_{\mathrm{Ry}} \ll \zeta$ (see Fig.~\ref{fig:lengthscales}). Our goal is to provide a description of the interaction and quantum dynamics of such a 1D string of Rydberg ions in a linear ion trap.

In comparison to neutral Rydberg atoms, a number of features and differences emerges for Rydberg ions in a Paul trap, which will be illuminated in the present work. First of all, the character of a Rydberg ion as a composite object gives rise to an intrinsic coupling of electronic and external motion in the presence of the electric trapping fields. This will be shown to result in renormalized trapping frequencies for Rydberg ions compared to their ground state counterparts.
Furthermore, the interaction among ions is not - as in the neutral case - governed by the dipole-dipole force alone but also by the charge-charge, dipole-charge and charge-quadrupole interaction. The interplay of this rich variety of interactions among the ions and the external trapping fields will be analyzed.

In our study we focus on ionic Rydberg states with low angular momentum quantum number and correspondingly large quantum defect, and large fine structure splitting, as these states can be most simply described and most easily excited by laser light from electronic ground states. For typical ion trap parameters \cite{Leibfried03} static dipolar and van-der-Waals interaction among ions in these Rydberg states is shown to be small compared to the energy scale set by the external trapping frequencies of the ions. Thus, in order to establish substantially stronger interactions, we employ additional microwave (MW) fields driving transitions between Rydberg states. This leads to large oscillating dipole moments, which result in remarkably strong controllable dipole-dipole interactions between the ions. Our findings show that for strong interactions the internal and external dynamics of the ion chain approximately decouple such that a ``frozen" Rydberg gas is formed. In this limit the Hamiltonian describing the electronic dynamics can be formulated in an effective spin-1/2 representation and involves resonant and off-resonant dipole-dipole interaction terms with coupling strengths of the order of several hundred MHz. Based on this effective spin model we demonstrate resonant excitation transfer on a nanosecond timescale from one end to the other of the ion chain. Moreover, we show that a two-qubit conditional phase gate between adjacent ions, based on the dipole-dipole interaction, can be realized on a time scale, which is much shorter than both the external dynamics and the radiative lifetime of the involved Rydberg states.

The paper is organized as follows: In Sec.~\ref{txt:single_rydberg_ions} we derive the Hamiltonian of the coupled internal and external dynamics of a single Rydberg ion in a linear Paul trap. We analyze the effects of the trapping fields onto the electronic properties of an ion excited to a Rydberg state, introduce MW-dressed Rydberg states and study the coupled electronic and external equations of motion within a Born-Oppenheimer approach. In Sec.~\ref{txt:interacting_ions} we turn to the analysis of several trapped Rydberg ions, discuss the variety of interactions among the ions and derive the effective spin-1/2 Hamiltonian for the electronic dynamics in the ion chain. Finally we illustrate the quantum dynamics of the system by studying resonant excitation transfer and present the fast two-qubit gate scheme.

\section{Single Trapped Rydberg Ions as a Composite Object}
\label{txt:single_rydberg_ions}

\subsection{Single Rydberg Ion in a Linear Paul Trap}
\label{txt:linear_paul_trap}
A combination of static and time-dependent electric fields can be employed to confine  charged particles in a restricted region of space. The electric potential of a quadrupole field of a standard Paul trap can be written in the form
\begin{eqnarray}
\Phi_T(\mathbf{r},t)=\alpha\cos\left(\omega
t\right)\left[x^2-y^2\right]-\beta\left[x^2+y^2-2z^2\right], \label{eq:paul_trap}
\end{eqnarray}
where $\omega$ is the radio frequency (RF) drive frequency. The electric field gradients  $\alpha$ and $\beta$
are determined by the actual geometry of the experimental setup. In present experiments, typical parameters are $\alpha\sim 10^9\,\mathrm{V}/\mathrm{m}^2$, $\beta\sim 10^7\,\mathrm{V}/\mathrm{m}^2$ and
$\omega=2\pi\times 10...100 \,\mathrm{MHz}$ (for details see e.g.~Ref.~\cite{Leibfried03}).

We are interested in the properties of Rydberg ions of alkali earth metals \cite{Djerad91} in an electric quadrupole trap. They possess a single valence electron with the remaining electrons forming closed shells \cite{Gallagher}. For the description of such system one can employ an effective two-body approach in which the ion is modeled by a two-fold positively charged core (mass $M$, position $\mathbf{r}_c$) and the valence electron (mass $m$, position $\mathbf{r}_e$). The corresponding interaction (model-)potential depends only on the relative coordinate $\mathbf{r}_c-\mathbf{r}_e$ and also on the angular momentum state of the atom. The latter dependence accounts for the quantum defect - a lowering in the energy for low angular momentum states in which the valence electron probes the inner electronic shells. High angular momentum states (typically with angular momentum quantum number $l>5$) do not exhibit a significant quantum defect since here the valence electron is located far away from the ionic core thus experiencing a bare Coulomb potential.

We now formulate the Hamiltonian of a single Rydberg ion in the presence of the electric potential of the Paul trap. We add a linear potential, corresponding to a time-dependent homogeneous electric field, $\Phi_l(\mathbf{r})=\mathbf{f}(t)\cdot\mathbf{r}$, such that the combined electric potential reads $\Phi(\mathbf{r},t)=\Phi_T(\mathbf{r},t)+\Phi_l(\mathbf{r},t)$. Below, we will employ these additional MW fields to electronically couple different Rydberg states. This will allow us to generate large oscillating dipole moments, which give rise to strong dipole-dipole interactions among dressed Rydberg ions.

Writing the interaction potential between the valence electron and the atomic core as $V(\left|\mathbf{r}_e-\mathbf{r}_c\right|)$ and taking into account the coupling of the individual charges to the electric potentials we find
\begin{eqnarray}
  H_\mathrm{lab}&=&\frac{\mathbf{p}_c^2}{2 M}  + \frac{\mathbf{p}_e^2}{2 m}
  +V(|\mathbf{r}_e-\mathbf{r}_c|)+2e\Phi(\mathbf{r}_c,t)-e\Phi(\mathbf{r}_e,t)+H_\mathrm{FS}.
\end{eqnarray}
The term $H_\mathrm{FS}$ accounts for the spin-orbit coupling giving rise to the fine-structure of electronic levels. Its effect will be discussed in the following subsection.

We introduce center of mass (CM) $\mathbf{R}$ and relative coordinates $\mathbf{r}$ according to
\begin{eqnarray}
  \mathbf{r}_c=\mathbf{R}-\frac{m}{M+m}\mathbf{r}\quad,\quad  \mathbf{r}_e=\mathbf{R}+\frac{M}{M+m}\mathbf{r}.
\end{eqnarray}
Exploiting that the nuclear mass is much larger than the electronic one and hence $M\gg m$ we have $\mathbf{r}_c\approx \mathbf{R}$ and $\mathbf{r}_e\approx \mathbf{R}+\mathbf{r}$. Within this approximation the Hamiltonian becomes
\begin{eqnarray}
  H&=&\frac{\mathbf{P}^2}{2 M}  + \frac{\mathbf{p}^2}{2 m}
  +V(|\mathbf{r}|)+2e\Phi(\mathbf{R},t)-e\Phi(\mathbf{R}+\mathbf{r},t)+H_\mathrm{FS}\\
  &=&\frac{\mathbf{P}^2}{2 M}  + \frac{\mathbf{p}^2}{2 m}
  +V(|\mathbf{r}|) \nonumber \\
  &&+e\left[\Phi(\mathbf{R},t)-\frac{\partial \Phi(\mathbf{R},t)}{\partial \mathbf{R}}\cdot \mathbf{r}-\frac{1}{2}\sum_{kl} x_k\frac{\partial^2\Phi(\mathbf{R},t)}{\partial X_k \partial X_l} x_l \right]+H_\mathrm{FS}.\nonumber\label{eq:single_ion_cm_rel}
\end{eqnarray}
Corrections to this Hamiltonian scale with $m/M$ which is typically about $10^{-5}$. 

In case of ions in low lying electronic states the potential (\ref{eq:paul_trap}) provides \textit{static} confinement along the longitudinal (z-)direction. However, transversally at no instant of time a confining potential is present. One rather finds a periodically oscillating potential saddle centered at the origin of the coordinate system. Due to the rapid periodic change of the confining and non-confining direction, however, the ions experience a ponderomotive potential that can provide transversal confinement \cite{Cook85}. In order to make this manifest we transform into a frame which oscillates at the RF frequency $\omega$ in the CM coordinate system. This is achieved by the unitary transformation
\begin{eqnarray}
  U=U(\mathbf{R},t) = \exp \left( - i \frac{e \alpha}{\hbar \omega} \left[ X^2 - Y^2 \right] \sin(\omega t)\right).
\end{eqnarray}
By applying this transformation to Hamiltonian (\ref{eq:single_ion_cm_rel}) one obtains
\begin{eqnarray}
H' & = & U H U^\dagger + i \hbar \frac{\partial U}{\partial t} U^\dagger = H_{\mathrm{CM}} + H_{\mathrm{el}} + H_{\mathrm{CM-el}} + H_{\mathrm{mm}}
\label{eq:SingleIon}
\end{eqnarray}
with
\begin{eqnarray}
H_{\mathrm{CM}} & = & \frac{\mathbf{P}^2}{2 M} + \frac{1}{2} M \omega_z^2 Z^2 + \frac{1}{2} M \omega^2_{\rho} \left( X^2 + Y^2 \right) \\
H_\mathrm{el} & = & \frac{\mathbf{p}^2}{2 m} + V\left( | \mathbf{r} |\right) - e \Phi (\mathbf{r},t)+ H_\mathrm{FS}
\label{eq:electronic_hamiltonian}\\
H_\mathrm{CM-el} & = & - 2 e \left[ \alpha \cos(\omega t) \left( Xx - Yy \right) - \beta \left( Xx + Yy - 2Zz \right) \right] \\
H_\mathrm{mm} & = & - \frac{2 e \alpha}{M \omega} \sin(\omega t) \left( X P_x - Y P_y \right)  \nonumber \\
& & - \frac{e^2 \alpha^2}{M \omega^2} (X^2 + Y^2) \cos(2\omega t) + e \mathbf{f}(t)\cdot \mathbf{R} .
\end{eqnarray}
Here $H_\mathrm{CM}$ provides harmonic axial and transversal confinement of the CM motion with the corresponding trap frequencies $\omega_z = 2 \sqrt{\frac{e \beta}{M}}$ and $\omega_\rho =  \sqrt{2} \sqrt{\left( \frac{e \alpha}{M \omega}\right)^2 - \frac{e \beta}{M}}$, which are of the order of a few MHz and satisfy $\omega_\rho \gg \omega_z$. For Ca$^+$ ions, an RF frequency $\omega = 2 \pi \times 15$ MHz and the gradient parameters $\alpha=10^9$ V/m$^2$ and $\beta=10^7$ V/m$^2$ the axial and radial trap frequencies evaluate to $\omega_z = 2 \pi \times 1.56 $ MHz and $\omega_\rho=2 \pi \times 5.64$ MHz, respectively.

The term $H_\mathrm{el}$ contains all dependencies on the electronic coordinates describing the motion of an electron in the field of a doubly charged ionic core which is superimposed by the electric potential $\Phi(\mathbf{r},t)$. The electronic dynamics takes place on a much faster time scale compared to the CM motion in the trap. The external electric field prevents, unlike in the field-free case, the separation of the CM and relative dynamics. The coupling between these motions is accounted for by $H_\mathrm{CM-el}$. Due to the large separation of time scales of electronic and external dynamics we will treat this coupling within an Born-Oppenheimer approach below.

Finally, we have the term $H_\mathrm{mm}$ which gives rise to the micromotion causing a coupling between the static oscillator levels of $H_{\mathrm{CM}}$. It can be shown \cite{Cook85} that for large enough values of the RF frequency $\omega$ this coupling can be neglected and the external motion of the ions can be considered as if it was taking place in a static harmonic potential. The effect of the additional micromotion term $e \mathbf{f}(t)\cdot \mathbf{R}$ arising from the MW dressing fields can likewise be neglected in the following, since typical MW frequencies are of the order of at least one GHz and therefore far from being resonant with the external trapping frequency of the ions.

With the full Hamiltonian (\ref{eq:SingleIon}) for the internal and external dynamics and the coupling among them at hand, we are now in the position to analyze the electronic properties of a trapped Rydberg ion and the mutual interplay of internal and external dynamics. This will be addressed in the next two subsections.

\subsection{Electronic Properties}
\label{txt:electronic_properties}
\begin{figure}\center
\includegraphics[angle=0,width=9cm]{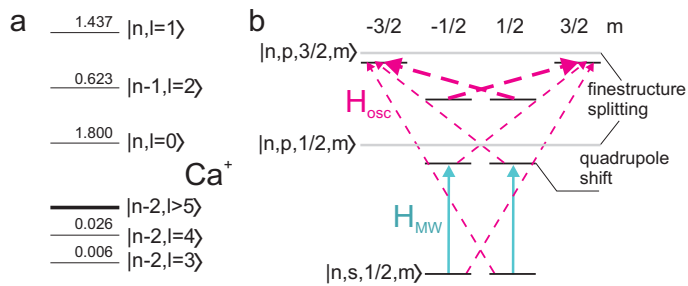}
\caption{\textbf{a}: Sketch of the Rydberg level structure of the $\mathrm{Ca}^+$ ion in the field free case without finestructure. States with $l>5$ do not exhibit a quantum defect and are located in a degenerate manifold of states. States with low angular momentum are split away from this manifold. The corresponding quantum defects are provided (taken from Ref. \cite{Djerad91}). \textbf{b}: Levels of the $s$ and $p$ manifold in the presence of the electric potential $\Phi(\mathbf{r},t)$. Its static components lead to a first order shift (quadrupole shift) of the energy levels according to eqs. (\ref{eq:energy-shifts1}) and (\ref{eq:energy-shifts2}). In general the electronic energy is lowered with respect to the field-free value (sketched by the solid gray lines). The shifted levels are coupled by the microwave Hamiltonian $H_\mathrm{MW}$ (blue, solid lines) and the oscillating components of the trapping field (red, dashed lines).
We consider the limit of large finestructure splitting in which the coupling between the states $\left|n,p,1/2,m\right>$ and $\left|n,p,3/2,m\right>$ (thin dashed lines), which is caused by $H_\mathrm{osc}$, can be neglected.
}\label{fig:spectrum}
\end{figure}

We proceed by inspecting the electronic Hamiltonian $H_\mathrm{el}$ in Eq.~(\ref{eq:electronic_hamiltonian}),
\begin{eqnarray}
H_\mathrm{el} & =  & \frac{\mathbf{p}^2}{2 m} + V\left( | \mathbf{r} |\right) +H_\mathrm{FS} +H_\mathrm{ef}
\end{eqnarray}
with the electric trapping and MW dressing fields contained in
\begin{eqnarray}
H_\mathrm{ef}&=&- e \Phi (\mathbf{r},t)= H_\mathrm{stat} + H_\mathrm{osc}+H_\mathrm{MW}\nonumber\\
&=&e\,\beta\left[x^2+y^2-2z^2\right]-e\,\alpha\cos\left(\omega
t\right)\left[x^2-y^2\right]-e\,\mathbf{f}(t)\cdot\mathbf{r}.
\label{eq:ion-field interaction}
\end{eqnarray}
In the absence of electric fields the ionic Rydberg states can be classified by the principal quantum number $n$, the angular momentum quantum number $l$, the total angular momentum $j$ and its magnetic quantum number $m$. The quantum states are represented by $\left|n,l,j,m\right>=\left|n,l\right>\left|j,m\right>_a$ which factors in the radial part $\left|n,l\right>$ and the angular momentum part $\left|j,m\right>_a$. The latter is constituted by a linear combination of products of the Spherical Harmonics and spin orbitals. The corresponding energies of the Rydberg levels are given by the well-known formula $E_{n\, l\, j}=-4\,E_\mathrm{Ryd}/(n-\delta(l))^2+E_\mathrm{FS}(n,l,j)$ where $E_\mathrm{Ryd}=13.6\,\mathrm{eV}$ is the Rydberg constant and $\delta(l)$ the quantum defect \cite{Gallagher,Djerad91}. The energy $E_\mathrm{FS}(n,l,j)$ accounts for the finestructure splitting due to $H_\mathrm{FS}$. The typical Rydberg level structure (without finestructure) is sketched in Fig.~\ref{fig:spectrum}a for the case of Ca$^+$.

In this paper we focus on states with large quantum defect, i.e. $s$ and $p$ states. This is motivated by the fact that these states are most easily accessible via laser excitations from the electronic ground states, and that they are energetically far separated from the degenerate manifold of states of higher angular momentum states. Generically, the energy separation $\triangle E_{l,l+1}$ between these states scales as $n^{-3}$ for large $n$. In case of $\mathrm{Ca}^+$ the energy separation between the $s$ and the $p$ level is $\triangle E_{s,p}/\hbar \sim 280$ GHz for $n=60$.

Let us now inspect the effect of the electron field interaction $H_\mathrm{ef}$ contained in the electronic Hamiltonian (\ref{eq:ion-field interaction}). We assume that the finestructure splitting of the $p$ state is sufficiently large such that neither the oscillating nor the static part of $H_\mathrm{ef}$ cause significant coupling between the states $\left|n,p,1/2,m\right>$ and $\left|n,p,3/2,m\right>$ and $j$ remains a good quantum number. Since the finestructure splitting scales proportional to $n^{-3}$ and becomes larger with increasing atomic mass the validity of this assumption can be ensured by either choosing not too high principal quantum numbers or by employing heavy ions. Moreover, we can neglect the coupling between the $s$ and $p$ states which is justified by their typical energy splitting of several 100 GHz. Considering first the static part $H_\mathrm{stat}$ of Eq.~(\ref{eq:ion-field interaction}) we find that it gives rise to the following quadrupole shifts which depend on $j$ and $m$:
\begin{eqnarray}
\label{eq:energy-shifts1}
  \triangle E_{s}&=&0\\
  \label{eq:energy-shifts2}
  \triangle E_{pj|m|}&=&\frac{2}{15}e\beta\left<n,p\mid r^2\mid n,p\right>\left[2|m|-j-\frac{9}{2}\right]. \label{eq:p_level_shift}
\end{eqnarray}
Here $\left<n,l\mid r^2\mid n,l\right>$ denotes the radial matrix element of $r^2$ calculated using the radial eigenfunctions that belong to the atomic interaction potential $V(|\mathbf{r}|)$. The total energy of the electronic states is hence given by $E_{n l j m}=E_{n l j}+\triangle E_{l j |m|}$. The quadrupole shifts with respect to the unperturbed energy of the $p$ states are sketched in Fig.~\ref{fig:spectrum}b.

For too strong electric field gradients there is the danger of ionizing the ion due to $H_\mathrm{stat}$. To estimate the ionization threshold we consider the potential
\begin{eqnarray}
  V^\prime=-\frac{2e^2}{4\pi\epsilon_0 r}+e\beta\left[x^2+y^2-2z^2\right].
\end{eqnarray}
Here we have approximated the interaction of the ionic core with the valence electron by a pure Coulomb potential. $V^\prime$ possesses two saddle points located at the $z$-axis at $z_\mathrm{sad}=\pm\left[\frac{e}{4\pi\epsilon_0\beta}\right]^{1/3}$ where it assumes the value $V^\prime_\mathrm{sad}=-\frac{3}{2}\left[\frac{e^5\beta}{\epsilon_0\pi^2}\right]^{1/3}$.
Solving $E_\mathrm{Ry}=V^\prime_\mathrm{sad}$ yields an estimate for the gradient $\beta$ at which field ionization would occur classically,
\begin{eqnarray}
  \beta_\mathrm{ion}=\frac{4}{27}\frac{e^7 m^3}{(4\pi \epsilon_0)^4 \hbar^6 n^6 }=1.44\times 10^{21} \frac{\mathrm{V}}{\mathrm{m}^2}\times \frac{1}{n^6}.
\end{eqnarray}
For $n=50$, for example, we find $\beta_\mathrm{ion}=9.2\times 10^{10}\, \mathrm{V}/\mathrm{m}^2$ which is one order of magnitude larger than the highest gradients actually used in experiment \cite{Leibfried03}. Hence, up to $n=50$ there is no danger for the Rydberg ion to undergo field ionization.

Let us now turn to the oscillating part of the electric field  which is accounted for by the term $H_\mathrm{osc}$ in the Hamiltonian (\ref{eq:ion-field interaction}). Since the oscillation frequency $\omega$ is typically $2\pi \times10...100 \,\mathrm{MHz}$, it is not sufficient to yield a resonant coupling between the $s$ and the $p$ states or between the fine structure components of the $p$ manifold. However, by estimating $|\left<n,p\mid r^2\mid n,p\right>|\approx a_0^2 n^4$, where $a_0$ is Bohr's radius, we find that $\left<H_\mathrm{osc}\right>\approx e\,\alpha\,a_0^2 n^4\cos(\omega\,t)$ and hence, although no resonant coupling is present, the strong modulation amplitude, which grows proportional to the fourth power of the principal quantum number $n$, might give rise to significant level shifts. From the symmetry properties of $H_\mathrm{osc}$ one concludes that only levels whose magnetic quantum number $m$ differs by 2 are coupled. These couplings are indicated in Fig.~\ref{fig:spectrum}b by the red, dashed arrows. Since we assume that the finestructure splitting is much larger than $|\left<H_\mathrm{osc}\right>|_\mathrm{max}$ the $j=1/2$ states are unaffected by $H_\mathrm{osc}$. By contrast strong effects are to be expected in the $j=3/2$ manifold.
However, in what follows we will exclusively work with the Rydberg states $\left|n,s\right>\equiv \left|n,s,1/2,m\right>$ and $\left|n,p\right>\equiv\left|n,p,1/2,m\right>$ (see Fig.~\ref{fig:spectrum}).

Finally, the term $H_\mathrm{MW}$ in Eq.~(\ref{eq:ion-field interaction}), which accounts for the additional MW dressing fields, is to be discussed. This will be subject of the following subsection.
\subsection{Microwave Dressing of Rydberg Levels}
\label{txt:dressed_rydberg_ions}
In this section we describe how MW dressing fields can be employed to generate strong interactions in an ion chain. As will be shown in Sec.~\ref{txt:interaction_between_rydberg_atoms} Rydberg ions aligned in a Wigner crystal in the Paul trap do not exhibit \textit{permanent} dipole moments, and residual Van-der-Waals interactions are small. Thus, our aim is to generate strong interactions using (near-)resonant MW dressing fields to couple electronic $s$ and $p$ Rydberg states as indicated in Fig.~\ref{fig:spectrum}b. We show in the following that the MW dressing gives rise to large \emph{oscillating} dipole moments, which in turn lead to strong dipole-dipole interaction between Rydberg ions.

The energy separation between $s$ and $p$ Rydberg states belonging to the same principal quantum number and thus as well the typical frequencies of the MW dressing fields are typically in the order of a few hundred GHz (cf.~Sec.~\ref{txt:electronic_properties}). Since the trapping frequencies determining the external motion of the ions are in the MHz range (cf.~Sec.~\ref{txt:linear_paul_trap}), coupling to the external motion is negligible and the MW fields exclusively affect the electronic degrees of freedom. Ideas similar to the MW dressing scheme described below were applied in the context of cold polar molecules, where a combination of static electric and MW
fields was used in order to tune intermolecular two- and three-body interactions \cite{Buechler07}.

\begin{figure}\center
\includegraphics[angle=0,width=7cm]{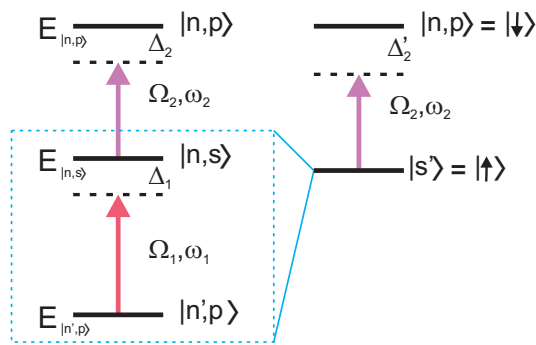}
\caption{Microwave dressing of ionic Rydberg states. A linearly polarized bichromatic microwave field is used to couple one $s$ with two $p$ levels one of which is far detuned. After an adiabatic elimination of the latter one obtains an effective two-level system. This dressing is used to tailor the interaction between Rydberg ions.}\label{fig:MW_dressing}
\end{figure}

In our setup we apply a linearly polarized MW such that non-zero transition dipole matrix elements occur between the states $\left|n,s\right>$ and $\left|n^\prime,p\right>$. For our purposes we choose a bichromatic microwave-field of the form $\mathbf{f}(t)=E_1\,\mathbf{e}_z\cos\omega_1 t+E_2\,\mathbf{e}_z\cos\omega_2 t$ such that the coupling Hamiltonian reads
\begin{eqnarray}
  H_\mathrm{MW}=-d_z\,\left[E_1 \cos\omega_1 t+E_2 \cos\omega_2 t\right]
\end{eqnarray}
where $d_z=e\,z$ is the operator of the dipole moment. We consider three levels (one $s$ level and two $p$ levels) which are coupled by $H_\mathrm{MW}$ as depicted in Fig.~\ref{fig:MW_dressing}. The frequencies $\omega_1$ and $\omega_2$ bridge the energy separations $E_{|n,s\rangle}-E_{|n',p\rangle}$ and $E_{|n,p\rangle}-E_{|n,s\rangle}$, respectively, and are assumed to differ significantly. This can be achieved by choosing not too close values of $n$ and $n^\prime$. In the rotating frame and after performing the rotating wave approximation the electronic Hamiltonian of a single ion reads
\begin{eqnarray}
H_\mathrm{3levels} & = & \hbar \triangle_1 |n^\prime,p\rangle \langle n^\prime,p| -\hbar \triangle_2 |n,p\rangle \langle n,p| \nonumber \\
& & + \frac{1}{2} \left[ \left(\hbar \Omega_1 |n,s\rangle \langle n^\prime,p| +\hbar \Omega_2 |n,p\rangle \langle n,s| \right) +\mathrm{h.c.} \right].
\label{eq:MW-Ham}
\end{eqnarray}
Here we have introduced the microwave detunings $\triangle_1 = \omega_1-(E_{|n,s\rangle}-E_{|n^\prime,p\rangle})/\hbar$ and $\triangle_2 = \omega_2-(E_{|n,p\rangle}-E_{|n,s\rangle})/\hbar$ and the Rabi frequencies $\Omega_{1,2}=-d_{1,2} E_{1,2} /3$ which involve the radial dipole matrix elements $d_1= e \langle n,s | r | n^\prime,p \rangle$ and $d_2 = e \langle n,s | r | n,p \rangle$. We assume that the MW field with frequency $\omega_1$ is far-detuned and only weakly couples the states $|n^\prime,p\rangle$ and $|n,s\rangle$. Furthermore it is assumed that $|\Delta_1| \gg |\Omega_2|, |\Delta_2|$. Under the condition $\Omega_1 \ll |\Delta_1|$ we can adiabatically eliminate the $|n^\prime,p\rangle$ state and obtain an effective two-level system consisting of the $|n,p\rangle$ state and a dressed state (see also Fig. \ref{fig:MW_dressing})
\begin{equation}
|s'\rangle = |n,s\rangle - \frac{\Omega_1}{2 \Delta_1} |n^\prime,p \rangle
\end{equation}
with $\eta=\Omega_1/(2 \Delta_1)$, $\eta\ll1$. This effective two-level system can be mapped onto a spin-1/2 particle by identifying the states $|s'\rangle$ and $|n,p\rangle$ as eigenstates of the spin operator $S_z$ with positive (negative) eigenvalue. The Hamiltonian (\ref{eq:MW-Ham}) then reduces to
\begin{equation}
H_0 = \frac{\hbar}{2}
\left( \begin{array}{cc} \Delta^\prime_2 & \Omega_2 \\
\Omega_2 & - \Delta^\prime_2 \end{array} \right) \equiv \mathbf{h} \mathbf{S}\label{eq:single_ion_MW}
\end{equation}
with an effective magnetic field $\mathbf{h}=(\Omega_2,0,\Delta^\prime_2)$, detuning $\Delta^\prime_2=\triangle_2-\Omega_1^2/(4\triangle_1)$ and the spin operator $\mathbf{S}=(S_x,S_y,S_z)$.

Due to the weak admixture of the state $|n^\prime,p\rangle$ the dressed state $|s'\rangle$ obtains an oscillating dipole moment. The matrix representation of the dipole operator $d_z$ in the set of states $|s'\rangle$, $|n,p\rangle$ is given by
\begin{equation}
d_z =\frac{1}{3}
\left(\begin{array}{cc}
-\frac{\Omega_1}{\Delta_1}d_1 \cos \omega_1 t &  d_2 e^{-i \omega_2 t} \\
 d_2e^{i \omega_2 t} & 0
\end{array}
\right).\label{eq:dipole_moment_MW}
\end{equation}
Using the abbreviation $D_{1,2}=(d_{1,2}/3)^2$ one obtains the following representation of the dipole moment operator:
\begin{eqnarray}
d_z = - \eta \sqrt{D_1} \cos(\omega_1 t) (1 + 2S_z) + 2\sqrt{D_2} (\cos (\omega_2 t) S_x + \sin(\omega_2 t) S_y).
\label{eq:dipole_moment_spin_notation}
\end{eqnarray}
The magnitude of the induced dipole moments is determined by the transition dipole matrix elements $d_{1,2}$ which can be roughly estimated as $d_{1,2} \sim e a_0 n^2$. Thus the dipole-dipole interaction energy scales $\sim n^4$ for large $n$. This is to be compared with the radiative life time of Rydberg states, which for large $n$ and low $l$ scales according to $\sim n^3$ \cite{Gallagher}, thereby favoring larger values of $n$. We return to the question of radiative decay and the validity of our analysis in Sec. \ref{txt:dressed_ion_interaction}, where we will use the representation (\ref{eq:dipole_moment_spin_notation}) of the dipole moment operator to derive an effective spin chain Hamiltonian describing the dynamics of Rydberg excitations in the ion chain.

\subsection{Coupling Between Internal and External Dynamics}
\label{txt:coupling_ext_int}
In this section we will analyze the effects of the coupling of internal and external degrees of freedom described by $H_\mathrm{CM-el}$ in the Hamiltonian (\ref{eq:SingleIon}). We treat these coupling terms using a Born-Oppenheimer approach, which is reasonable since the electronic dynamics takes place on a much faster time scale than the external motion of the ions. In this framework we treat the CM coordinates as parameters while diagonalizing the electronic Hamiltonian. We evaluate the energy of the $|n,s\rangle$-level by second order perturbation theory considering only the coupling to the next $|n,p\rangle$-level. After averaging over one RF cycle one obtains the energy correction
\begin{eqnarray}
  \epsilon(X,Y,Z)=E_{|n,s\rangle}+\frac{1}{2}M{\omega_z^\prime}^2 Z^2+\frac{1}{2}M{\omega_\rho^\prime}^2(X^2+Y^2)
\end{eqnarray}
with
\begin{eqnarray}
{\omega_z^\prime}^2&=&-\frac{32}{3M}\frac{e^2\beta^2}{\triangle E_{s,p}}\left|\left<n,s\mid r \mid n,p\right>\right|^2,\\
{\omega_\rho^\prime}^2&=&-\frac{2}{3M}\frac{e^2(\alpha^2+2\beta^2)}{\triangle E_{s,p}}\left|\left<n,s\mid r \mid n,p\right>\right|^2,
\end{eqnarray}
and the energy difference $\triangle E_{s,p} = E_{|n,p\rangle}-E_{|n,s\rangle}$. Hence, ions excited to Rydberg states experience modified transversal and longitudinal trap frequencies in comparison to their ground state counterparts. For the $s$ state under consideration the trap becomes shallower and the new trap frequencies are given by $\tilde{\omega}_{\rho,z}=\omega_{\rho,z}\sqrt{1+(\omega^\prime_{\rho,z}/\omega_{\rho,z})^2} \approx \omega_{\rho,z} + \omega^{\prime 2}_{\rho,z}/(2\omega_{\rho,z})=\omega_{\rho,z}+\delta \omega_{\rho,z}$, yielding
\begin{eqnarray}
\frac{\delta \omega_z}{\omega_z} & = & \frac{4}{3} e \beta \frac{\left|\left<n,s\mid r \mid n,p\right>_r\right|^2}{\triangle E_{s,p}}, \\
\frac{\delta \omega_\rho}{\omega_\rho} & = & \frac{1}{6} \frac{\alpha^2+2 \beta^2}{\alpha^2/(M\omega^2) -\beta/e}  \frac{\left|\left<n,s\mid r \mid n,p\right>_r\right|^2}{\triangle E_{s,p}}.
\end{eqnarray}
By estimating $\left|\left<n,s\mid r \mid n,p\right>\right| \sim a_0 n^2$ one finds that the frequency shift $\delta \omega_{\rho,z}$ scales proportional to $n^7$. For typical trap parameters, i.e.~$\alpha \gg \beta$, the modification of the axial trap frequency is much smaller than the corresponding one for the radial frequency. For $n=50$ and the parameters presented in Sec.~\ref{txt:electronic_properties} we find $(\delta \omega_\rho) /\omega_\rho = 3.5 \times 10^{-2} $ and $(\delta \omega_z) /\omega_z = 7.4 \times10^{-4}$. Below we are interested in the regime where these modifications of the trapping frequencies are negligible. We remark, though, that due to the scaling $\sim n^7$ they can become significant even for moderately larger principle quantum numbers.

\section{Interacting Trapped Rydberg Ions}
\label{txt:interacting_ions}
We now turn to the discussion of the interaction between several ions. First, we analyze the various types of emerging interactions between Rydberg ions, and discuss the interplay of interionic interactions and the effect of the external trapping fields. Then we proceed with the discussion of interacting Rydberg ions dressed by MW fields and derive the corresponding effective spin-1/2 Hamiltonian describing the Rydberg excitation dynamics in the ion chain. The role of the two effective spin states of each ion will be played by two Rydberg states, as outlined in Sec.~\ref{txt:dressed_rydberg_ions} and illustrated in Fig.~\ref{fig:MW_dressing}.

We illustrate the effective spin dynamics by studying the process of resonant excitation transfer in a mesoscopic chain of ten ions. Furthermore, we suggest a scheme, which allows to observe the effective spin dynamics in dressed ground state ions. We finally show how the trapped Rydberg ion system can be exploited in the context of quantum information processing and discuss an implementation of a two-qubit conditional phase gate.
\subsection{Static Interactions between Trapped Rydberg Ions}
\label{txt:interaction_between_rydberg_atoms}
\begin{figure}\center
\includegraphics[angle=0,width=6cm]{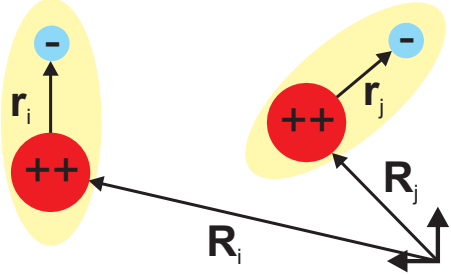}
\caption{Interacting Rydberg ions. The net charge of the ions leads, apart from the common Coulomb repulsion, to a charge-dipole and a charge-quadrupole interaction both of which are absent in systems of neutral Rydberg atoms.}\label{fig:two_ions}
\end{figure}

We consider the interaction between ions $i$ and $j$ with coordinates given by $(\mathbf{R}_i, \mathbf{r}_i)$ and $(\mathbf{R}_j, \mathbf{r}_j)$ as depicted in Fig.~\ref{fig:two_ions}.
The Coulomb interaction $V(\mathbf{r}_i, \mathbf{r}_j,\mathbf{R}_i,\mathbf{R}_j)$ between the charges belonging to different ions can be written as
\begin{eqnarray}
  \frac{V(\mathbf{R}_i, \mathbf{R}_j,\mathbf{r}_i,\mathbf{r}_j)}{e^2/(4 \pi \epsilon_0)} & = & \frac{4}{|\mathbf{R}_i - \mathbf{R}_j|} - \frac{2}{|\mathbf{R}_i-(\mathbf{R}_j + \mathbf{r}_j)|} \nonumber \\
  & & - \frac{2}{|(\mathbf{R}_i+\mathbf{r}_i) - \mathbf{R}_j|} + \frac{1}{|(\mathbf{R}_i+\mathbf{r}_i) - (\mathbf{R}_j+\mathbf{r}_j)|}.
\label{eq:multipole_expansion}
\end{eqnarray}
We assume that $|\mathbf{R}_i - \mathbf{R}_j|\gg |\mathbf{r}_{i/j}|$ which is very well fulfilled, since - as discussed above - we are interested in the parameter regime where the average interparticle distance $\zeta$ in the ion trap is much larger than the extension of the electronic wave function $a_\mathrm{Ry}$ of the Rydberg ions. Performing a multipole expansion up to second order in the small parameter $(a_\mathrm{Ry}/\zeta)$ and abbreviating $|\mathbf{R}_i-\mathbf{R}_j|=|\mathbf{R}_{ij}|=R_{ij}$ and $\mathbf{n}_{ij}=(\mathbf{R}_i-\mathbf{R}_j)/R_{ij}$ we obtain the following form for the Rydberg ion-Rydberg ion interaction potential:
\begin{eqnarray}
\frac{V(\mathbf{R}_i,\mathbf{R}_j,\mathbf{r}_i,\mathbf{r}_j)}{e^2/(4\pi \epsilon_0)}&  = & \frac{1}{R_{ij}} + \frac{(\mathbf{R}_i-\mathbf{R}_j) (\mathbf{r}_i -\mathbf{r}_j)}{R_{ij}^3}\nonumber\\
 &&+ \frac{r_i^2 -3(\mathbf{n}_{ij}\cdot \mathbf{r}_i)^2 + r_j^2 -3(\mathbf{n}_{ij}\cdot \mathbf{r}_j)^2}{2 R_{ij}^3} \nonumber \\
& & + \frac{\mathbf{r}_i\cdot \mathbf{r}_j-3 (\mathbf{n}_{ij}\cdot \mathbf{r}_i) (\mathbf{n}_{ij}\cdot \mathbf{r}_j)}{R_{ij}^3}.
\end{eqnarray}
The first term accounts for the Coulomb interaction between two singly charged ions and is independent of the degree of electronic excitation. In case of Rydberg ions the displacement of the electronic charge from the ionic core leads to the exhibition of a dipole moment which interacts with the charge of the other ion. This dipole-charge interaction gives rise to the second term. The third term accounts for the charge-quadrupole interaction. These three terms are absent in the case of interacting neutral Rydberg atoms. The last term is the well-known dipole-dipole interaction, which is also present in neutral systems.

For $N$ ions stored in a Paul trap, at sufficiently low temperature and tight radial trapping, $\omega_\rho \gg \omega_z$, the ions form a one-dimensional Wigner crystal with equilibrium positions along the $Z$-axis \cite{James98,Porras04}, which are determined by the interplay between the Coulomb repulsion among the ions and the external confinement by the trapping fields. In \ref{app:interacting_ions} we show that after a harmonic expansion of the Hamiltonian around the equilibrium positions $Z_i^{(0)}$ (given by Eq.~(\ref{eq:equ_condition})) the full Hamiltonian of $N$ interacting ions can be written as
\begin{eqnarray}
H_\mathrm{ions} & = & H_\mathrm{ph}+ \sum_i^N H_{\mathrm{el},i} + H_\mathrm{int-ext} + H_\mathrm{dd}.
\label{eq:N_ion_Hamiltonian}
\end{eqnarray}
The first term
\begin{equation}
\label{eq:full_N_ion_Hamiltonian}
H_\mathrm{ph}=\sum_{\alpha=x,y,z} \sum_{n}^N \hbar \omega_{\alpha,n} a_{\alpha,n}^\dagger a_{\alpha,n}
\end{equation}
describes the external oscillation dynamics of the ionic cores with $a_{\alpha,n}^\dagger$ and $a_{\alpha,n}$ being the respective creation and annihilation operators of the normal modes (cf.~Eqs.~(\ref{txt:H_ph_1_appendix})-(\ref{eq:M_matrices_appendix})). The second term determines the electronic level structure of the ions: the charge-quadrupole term gives rise to a position dependent variation of the electric field experienced by the trapped ions, which can be absorbed in the single particle ion-field interaction $H_\mathrm{ef}$ given by Eq.~(\ref{eq:ion-field interaction}). Thus, the electronic Hamiltonian of the $i$-th ion takes the form
\begin{eqnarray}
  H_\mathrm{el,i}&=&\frac{\mathbf{p}_i^2}{2 m} + V\left( | \mathbf{r}_i |\right) + e\,\beta^\prime_i\left[x_i^2+y_i^2-2z_i^2\right]-e\,\alpha\cos\left(\omega
t\right)\left[x_i^2-y_i^2\right]\nonumber\\
&&-e\,\mathbf{f}(t)\cdot\mathbf{r}_i\label{eq:internal_Ham_including_quad}
\end{eqnarray}
with position dependent gradient
\begin{eqnarray}
\beta^\prime_i=\beta+\frac{e}{8 \pi \epsilon_0} \sum_{j(\neq i)}^N \frac{1}{| Z_i^{(0)} - Z_j^{(0)}|^3}=\beta+\delta\beta_i.
\label{eq:position_dependent_gradient}
\end{eqnarray}
The third term $H_\mathrm{int-ext}$ in Eq.~(\ref{eq:N_ion_Hamiltonian}) accounts for the coupling of the internal and external dynamics, which is partly due to the dipole-charge interaction and partly due to the inhomogeneous electric field of the Paul trap (see Eq.~(\ref{eq:int_ext_coupling})). The resulting coupling between the internal and external dynamics becomes also position dependent and thus leads to a state-dependent variation of the trapping frequency (see Sec.~\ref{txt:coupling_ext_int}). However, since we work in a regime where these shifts are negligible already in the single-ion case they can also be neglected in the case of many ions.

Finally we have the dipole-dipole interaction in Eq.~(\ref{eq:N_ion_Hamiltonian}), which after the harmonic expansion is given by
\begin{eqnarray}
H_\mathrm{dd}=\frac{1}{2}\frac{e^2}{4 \pi \epsilon_0} \sum_{i\neq j}^N \frac{\mathbf{r}_i \cdot \mathbf{r}_j - 3 z_i z_j}{| Z_i^{(0)} - Z_j^{(0)}|^3}.
\end{eqnarray}

The discussion of the individual terms contained in Hamiltonian (\ref{eq:N_ion_Hamiltonian}) proceeds in analogy to the one presented for the single ion case. The electronic dynamics of each ion is governed by $H_{\mathrm{el},i}$ in Eq.~(\ref{eq:N_ion_Hamiltonian}) and is therefore dependent on the equilibrium positions of the ions. This gives rise to position dependent electronic energies, i.e.~$E_{nljm} \rightarrow E_{{nljm},i}$ according to Eqs.~(\ref{eq:ion-field interaction})-(\ref{eq:energy-shifts2}) with the ion-dependent gradients of Eq.~(\ref{eq:position_dependent_gradient}).

We remark that even ions located at the edges of the ion chain experience the same electric field as an ion in the center of the trap. This is due to the fact that the electric field of the Paul trap is compensated by the field created by the ions themselves. Therefore, the ions do not exhibit permanent dipole moments and each ion can be considered to be located in a center of a local electric quadrupole field. This is the reason why the electronic properties of the system of trapped ions are virtually unaffected by the actual number of particles at hand.

The term of interest to create strong interactions among the ions is the dipole-dipole interaction. In the parameter regime where the interparticle distance $\zeta$ is much larger than the extension of the electronic wave function $a_\mathrm{Ry}$ of a Rydberg ion, the dipole-dipole interaction can be treated perturbatively. For the considered $s$ and $p$ states the expected interaction energy can be estimated by $E_\mathrm{vdW} \sim e^4|\left<n,s\mid r \mid n,p\right>|^4/((4\pi\epsilon_0)^2|\triangle E_{s,p}| \zeta^6)$. For $\mathrm{Ca}^+$, $n=50$ and $\zeta=5\,\mu\mathrm{m}$ one finds $E_\mathrm{vdW}\sim \hbar\times 200\,\mathrm{kHz}$. In the next subsection we demonstrate that the MW fields introduced in Sec.~\ref{txt:dressed_rydberg_ions} in order to manipulate the electronic level structure can give rise to much stronger interactions among the ions.

\subsection{Interaction Between MW-Dressed Rydberg Ions}
\label{txt:dressed_ion_interaction}
To study the interaction between MW-dressed Rydberg ions we first transform the $N$-ion Hamiltonian (\ref{eq:N_ion_Hamiltonian}) to a frame of reference, which rotates at the microwave frequencies $\omega_1$ and $\omega_2$ of the dressing fields (in analogy to the single-ion transformation in Sec.~\ref{txt:dressed_rydberg_ions}). In the rotating frame, the terms in $H_{\mathrm{int-ext}}$ oscillate rapidly at the frequencies $\omega_1$ and $\omega_2$ and $\omega_1-\omega_2$. Hence, in the rotating-wave approximation the internal and external dynamics decouple (cf.~discussion in Sec.~\ref{txt:linear_paul_trap}).

We use the dressed ion Hamiltonian (\ref{eq:single_ion_MW}) and the expression (\ref{eq:dipole_moment_MW}) for the electronic dipole moment in order to represent the Hamiltonian $H_\mathrm{int}=\sum_i^N H_{\mathrm{el},i}+ H_\mathrm{dd}$ which is part of the full Hamiltonian (\ref{eq:N_ion_Hamiltonian}). The terms $H_{\mathrm{el},i}$ are substituted by the two-dimensional representation given by Eq.~(\ref{eq:single_ion_MW}). In this representation we can write the dipole-dipole interaction as
\begin{eqnarray}
  H_\mathrm{dd}=-\frac{1}{4 \pi \epsilon_0} \sum_{i\neq j}^N \frac{d_{z}^{(i)}d_{z}^{(j)}}{| Z_i^{(0)} - Z_j^{(0)}|^3}
\end{eqnarray}
where $d_{z}^{(i)}$ is given by Eq.~(\ref{eq:dipole_moment_MW}) with the index $i$ labeling the respective ion. In the rotating wave approximation we obtain
\begin{eqnarray}
\label{eq:spin_chain_Hamiltonian}
H_\mathrm{int}&=&\sum_i^N \mathbf{h}^{(i)} \mathbf{S}^{(i)} + D_1\sum_{i,j(\neq i)}^N\nu_{ij} \left[ \frac{1}{4}\eta_i^2  + \eta_i \eta_j S_z^{(i)}\right]\\
&&+ \sum_{i,j(\neq i)}^N \nu_{ij} \left[ D_2 \left( S_x^{(i)} S_x^{(j)} + S_y^{(i)} S_y^{(j)} \right) + \eta_i \eta_j D_1S_z^{(i)} S_z^{(j)} \right]\nonumber,
\end{eqnarray}
with $\nu_{ij} = -2/(4 \pi \epsilon_0 |Z_i^{(0)}-Z_j^{(0)}|^3)$ and an effective magnetic field $\mathbf{h}^{(i)}=\hbar(\Omega_2,0,\Delta_2^{(i)})$. The coefficients $D_{1,2}$ and $\eta_i$ characterize the coupling strengths and depend on the transition matrix elements between the involved Rydberg states and the MW dressing. As depicted in Fig.~\ref{fig:MW_dressing} the effective spin corresponds to a two-level system constituted by two MW-dressed Rydberg states (see Sec.~\ref{txt:dressed_rydberg_ions} for details).

\begin{figure}\center
\includegraphics[angle=0,width=7cm]{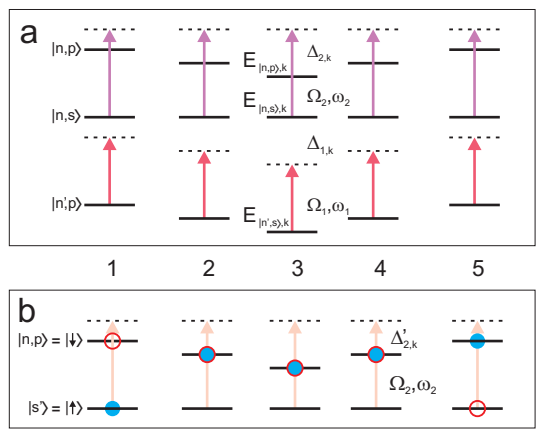}
\caption{\textbf{a}: Schematic level scheme of the three MW-coupled Rydberg levels, for a chain of 5 ions. The individual, position-dependent energy shifts of the $l=1$ states give rise to an inhomogeneous distribution of MW detunings (not true to scale). \textbf{b}: Schematic spin dynamics of a chain with the initial configuration: first ion in the $|n,p\rangle$ state, all other ions in the $|s'\rangle$ state (blue filled circles). After a certain time the Rydberg excitation has travelled to the right end of the chain (red open circles). A numerical example for this excitation transfer is shown in Fig.~\ref{fig:excitation_transfer}.}\label{fig:ion_chain}
\end{figure}

The terms linear in the spin operators represent a coupling of a series of spins to an inhomogeneous effective magnetic field whose strength and direction are determined by the position dependent electronic energies of the Hamiltonian (\ref{eq:internal_Ham_including_quad}). The terms in the second line of Eq.~(\ref{eq:spin_chain_Hamiltonian}) (quadratic in the spin operators) represent a ferromagnetic Heisenberg chain with $1/r^3$ exchange-type interaction \cite{Auerbach}. Establishing the connection to neutral Rydberg gases the $S_x^{(i)} S_x^{(j)}$ and $S_y^{(i)} S_y^{(j)}$ can be interpreted as resonant dipole-dipole interaction terms. The term which is proportional to $S_z^{(i)} S_z^{(j)}$ resembles the interaction of two static dipoles. In general, the coefficients $\eta_i=\Omega_1^{(i)}/(2 \Delta_1^{(i)})$ depend on the ion index, since different energy shifts of the $|n',p\rangle$ level due to position-dependent charge-quadrupole terms (cf.~Eqs.~(\ref{eq:internal_Ham_including_quad}) and (\ref{eq:position_dependent_gradient})) lead to different detunings $\Delta_1^{(i)}$ for different ions (see Fig.~\ref{fig:ion_chain}). In case one does not admix the $|n',p\rangle$ state to $|n,s\rangle$ (i.e.~$\Omega_1=0$ in Eq.~(\ref{eq:MW-Ham})), the coefficients $\eta_i$ vanish and the interaction Hamiltonian solely describes resonant dipole-dipole interaction in the presence of an effective magnetic field with a constant component in $x$ and a position-dependent component in $z$-direction.

The physical realization of effective spin dynamics, as provided by the Hamiltonian (\ref{eq:spin_chain_Hamiltonian}), has been of significant interest in atomic physics during the last few years as ``analog quantum simulators'' of (mesoscopic) condensed matter systems. The distinguishing feature of the present setup is the large coupling strength between effective spins, which scales proportional to $n^4$ and is of the order $\hbar\times 500\,\mathrm{MHz}$ ($n \approx 50$, typical interparticle spacing $\zeta \approx 5\,\mu\mathrm{m}$). The characteristic time scale of the effective spin dynamics is thus of the order of a few nanoseconds. This is significantly shorter than the typical decoherence time in the system, which is set by the radiative lifetimes of the involved Rydberg states, which scale as $~n^3$ and are  typically of the order of $\mu$s (for Ca$^+$ ions and $n=50$ the lifetime is $\sim 10 \,\mu$s \cite{Djerad91}).  

Realization of effective spin models has also been proposed in the context of trapped ions in their electronic ground state, and with cold atoms and polar molecules in optical lattices. For trapped ions involving electronic ground states models analogous to (\ref{eq:spin_chain_Hamiltonian}) can be derived where the  typical coupling strengths for the effective spin-spin interactions are in the range of tens of kHz \cite{Porras04} with (long) decoherence times as described in the context of ion trap quantum computing  \cite{Calarco01,Friedenauer08}.
Effective Heisenberg models with nearest neighbor interactions are also obtained with cold atoms in optical lattices \cite{Bloch07,Lewenstein07}, where the time scales of exchange interactions can be of the order of a few hundred Hz \cite{Anderlini07,Trotzky08}. We note that these energy scales are also directly related to the temperature requirements for the preparation of an effective zero temperature ensemble. Finally, effective spin models, such as the Kitaev model \cite{Kitaev06}, have been proposed with polar molecules in optical lattices \cite{Micheli06,Brennen07,ColdPolarMolecules}. In this context electric dipole moments of a few Debye can be induced by external DC and microwave electric fields in the rotational manifold of molecules prepared in their electronic and vibrational ground state. The electric dipole-dipole interactions, which are strong and long range in comparison with neutral atom collisional interactions, can lead to effective offsite-coupling strengths of the order of tens or hundred kHz, limited mainly by the conditions imposed by optical trapping, while coherence times are of the order of seconds, as determined e.g.~by spontaneous emission in off-resonant light fields forming the trapping potential. 

A second feature of simulating spin dynamics according to Hamiltonians of the type (\ref{eq:spin_chain_Hamiltonian}) is the single ion addressability and read out \cite{Roos04} which the present setup inherits from the experimental developments in ion trap quantum computing. In contrast, neutral atoms and molecules in optical molecules ususally allow global addressing by laser light, even though a significant effort is devoted at the moment to develop these tools also for optical lattice setups  \cite{Nelson07,Gorshkov08}. Note, however, that neutral atoms and molecules in optical lattices will typically allow for systems with a significantly larger number of ``spins'' than in the ion case.

One of the experimentally most challenging aspects of realizing a Rydberg ion chain is the requirement of $\pi$-pulses to transfer ions from the ground state to the Rydberg state. In Sec.~\ref{txt:adiabatic_scheme} we discuss a version of an effective spin chain where the Rydberg dipoles are admixed to the electronic ground states with an off-resonant laser process, resulting in ground state ions with effective oscillating dipole moments.

\begin{figure}\center
\includegraphics[angle=0,width=7cm]{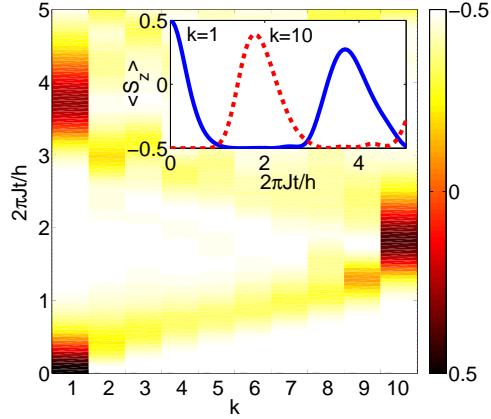}
\caption{Transport of an excitation through a chain of ten ions. The ion at site 1 is initially excited to the state $\left|\uparrow\right>$ while all others are in the state $\left|\downarrow\right>$. After the time $t=1.8\,\hbar/J$ the excitation is transferred from the first to the tenth ion. The inset shows the time evolution of the expectation values $\langle S^{(1)}_z\rangle$ and $\langle S^{(10)}_z\rangle$.}\label{fig:excitation_transfer}
\end{figure}

\subsection{Resonant Excitation Transfer}
\label{txt:resonant_excitation_transfer}
As an illustration of the spin dynamics contained in Hamiltonian (\ref{eq:spin_chain_Hamiltonian}), we study the transfer of an excitation from one side of the Rydberg ion chain to  the other end. For simplicity we choose a scenario in which $\eta_i=0$ ($\Omega_1=0$ in Eq.~(\ref{eq:MW-Ham})). As discussed in the previous subsection in this case the interaction reduces to purely resonant dipole-dipole interaction. The ion-dependent coefficient of the dipole-dipole interaction can be written as
\begin{eqnarray}
  \nu_{ij} = -\frac{2}{4 \pi \epsilon_0 |Z_i^{(0)}-Z_j^{(0)}|^3}=- \frac{2M\omega_z^2}{e^2}\frac{1}{|u_i-u_j|^3}
\end{eqnarray}
where the $u_i$ are the equilibrium positions in dimensionless coordinates, $u_i=Z_i^{(0)}/\zeta$ with $\zeta=\left[e^2/(4\pi\epsilon_0 M \omega_z^2)\right]^{1/3}$ (see Ref.~\cite{James98}). The scale of the interaction energy is then given by $J=- \frac{2M\omega_z^2}{e^2} D_2$. We choose the MW frequency $\omega_2$ such that it is on resonance with the energy gap between the levels $\left|n,s\right>$ and $\left|n,p\right>$ determined by the gradient (\ref{eq:position_dependent_gradient}) with $\delta\beta_i=0$. Following Eq.~(\ref{eq:p_level_shift}), the position dependent change of the gradient $\delta\beta_i$ gives rise to a position dependent variation of the detuning which reads
\begin{eqnarray}
\triangle_{2,i}&=&-\frac{4}{5}e\,\delta\beta_i\left<n,p\mid r^2 \mid n,p\right>\nonumber\\
&=&-\frac{2}{5} M \omega^2_z \left<n,p\mid r^2 \mid n,p\right> \sum_{j\neq i}^N \frac{1}{|u_i-u_j|^3}=B_z \sum_{j\neq i}^N \frac{1}{|u_i-u_j|^3}.
\end{eqnarray}
This situation is depicted in Fig.~\ref{fig:ion_chain}. The effective magnetic field $B_z$ and the coupling constant $J$ do not scale independently since both $D_2/e^2$ and the matrix element $\left<n,p\mid r^2 \mid n,p\right>$ are of the order of  $a^2_0 n^4$. The precise value depends on the ionic species. For our simulations we use the parameters $B_z=0.65\,J$ and $\hbar\Omega_2=0.01\,J$. Initially the system is prepared such that the first ion ($k=1$) is in the state $\left|\uparrow\right>=\left|n,s\right>$ while all others are in the state $\left|\downarrow\right> = \left| n,p\right> $. This state is sketched in Fig.~\ref{fig:ion_chain}b by the solid circles. The temporal evolution under the Hamiltonian (\ref{eq:spin_chain_Hamiltonian}) leads to a transfer of the excitation from the first ion to the last ion in the chain (open circles in Fig.~\ref{fig:ion_chain}b). The corresponding numerical data is shown in Fig.~\ref{fig:excitation_transfer} where we monitor the time evolution of the expectation value $\langle S_z^{(i)}\rangle$ in a chain of 10 ions. The excitation transfer from the first to the tenth ion takes place in a time $t=1.8\hbar/J$ with an efficiency of $89\,\%$. Since $J$ can be in the order of $\hbar\times 500\,\mathrm{MHz}$ this resonant excitation transfer can therefore be achieved in about $3.6\,\mathrm{ns}$ which is much less than the lifetime of ionic Rydberg states.

\subsection{Spin Dynamics in the Electronic Ground States}
\label{txt:adiabatic_scheme}
In view of the short transition wave length of $100...125 \,$nm associated to a transition from an electronic ground state to a Rydberg level \cite{NIST-Tables}, it is experimentally challenging to realize $\pi$ laser pulses, which transfer the entire electronic population to the Rydberg levels, as required for the initialization step of the chain of effective spins. Thus, we outline an alternative scheme, which does not require the transfer of the full electronic population to Rydberg states and which is based on an adiabatic admixture of Rydberg levels to two electronic ground states by near-resonant CW lasers.

\begin{figure}\center
\includegraphics[angle=0,width=6cm]{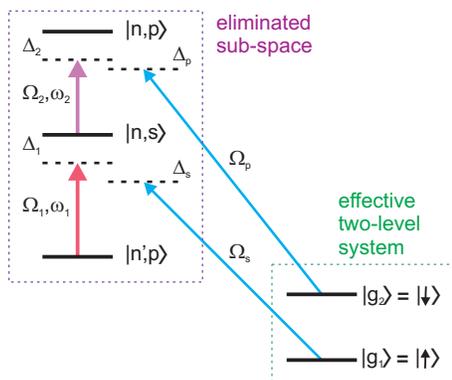}
\caption{MW dressing of ground state ions. Two lasers with Rabi frequencies $\Omega_s$, $\Omega_p$ and detunings $\Delta_s$, $\Delta_p$ weakly couple two electronic ground states $|g_1\rangle$, $|g_2\rangle$ to the Rydberg levels $|n,s\rangle$ and $|n,p\rangle$, respectively. Thus ions residing in the electronic ground states are dressed and obtain time-dependent oscillating dipole moments.}
\label{fig:MW_dressing_ground_states}
\end{figure}

We extend the 3-level scheme of Sec.~\ref{txt:dressed_rydberg_ions} by including two ground states $|g_1\rangle$ and $|g_2\rangle$, which are coupled by two lasers to the (undressed) Rydberg states $|n,s\rangle$ and $|n,p\rangle$, respectively (see Fig.~\ref{fig:MW_dressing_ground_states}). As before, we use the MW dressing fields to couple the set of bare Rydberg states $|n',p\rangle$, $|n,s\rangle$ and $|n,p\rangle$. The two CW lasers weakly couple the two ground states to the resulting dressed Rydberg level structure. Weak coupling requires that the laser Rabi frequencies are small compared to the MW Rabi frequencies and that the laser fields are sufficiently far detuned from the three dressed Rydberg states. Due to the laser coupling the two dressed ground states $|g_1'\rangle$, $|g_2'\rangle$ obtain time-dependent oscillating dipole moments, which finally lead to dipole-dipole interactions between laser-dressed \emph{ground state} ions. The dressed ground states now constitute the two-level system of interest (effective spin degree of freedom) and play the role of the dressed Rydberg levels $|s'\rangle$ and $|n,p\rangle$ of Sec.~\ref{txt:dressed_rydberg_ions}.

As shown in \ref{txt:dipole_moment_ground_states} for a specific set of Rabi frequencies and detunings of the MW and laser fields the representation of the dipole operator in this effective two-level system reads
\begin{eqnarray}
d_z & = & \left(\begin{array}{cc}
0 & c_{p,1} c_{s,2} \sqrt{D_2} e^{-i \omega_2 t} \\
c_{p,1} c_{s,2} \sqrt{D_2} e^{i \omega_2 t} & 2 c_{s,2} c_{p',2} \sqrt{D_1} \cos(\omega_1 t)\label{eq:dipole_operator_ground_states}
\end{array}
\right) \nonumber \\
& = & c_{s,2} c_{p',2} \sqrt{D_1} \cos(\omega_1 t) (1 - 2S_z) \nonumber \\
& & + 2 c_{p,1} c_{s,2} \sqrt{D_2} (\cos(\omega_2 t)S_x +\sin(\omega_2 t) S_y)
\end{eqnarray}
with the factors $c_\alpha$ given by Eqs.~(\ref{eq:admixture_coefficients}). Thus, the resulting dipole operator as well as the resulting spin chain Hamiltonian describing the dynamics in the dressed ground states $|g_1'\rangle$, $|g_2'\rangle$ have the same structure as in Eqs.~(\ref{eq:dipole_moment_spin_notation}) and (\ref{eq:spin_chain_Hamiltonian}), with the difference that the magnitude of the dipole moments is decreased by the factors $c_{p,1} c_{s,2}$ and $c_{p',2} c_{s,2}$, respectively.

Due to the weak admixture of the Rydberg states the dressed ground states have a finite lifetime, i.e. the time until a photon of the dressing lasers is scattered. Denoting the lifetime of the involved Rydberg states by $\tau=1/\Gamma$, this scattering rate is approximately given by $\Gamma_\mathrm{scat}=|c_\alpha|^{2}\Gamma$, where $c_\alpha$ represents a typical value of the admixture coefficients in Eq.~(\ref{eq:admixture_coefficients}). For instance, the lifetime of the state $|g_1'\rangle$ is enhanced by a factor of $|c_{p,1}|^{-2}$ with respect to the radiative lifetime of the Rydberg state $|n,p\rangle$, since only the fraction $|c_{p,1}|^2$ of the electronic population resides in the Rydberg state. Moreover, the fraction of the admixed Rydberg states affects the interaction among the ions. Compared to the bare dipole-dipole interaction the interaction between the dressed ions is suppressed by a factor $\sim |c_\alpha|^4$ resulting in an effectively slower spin dynamics. The typical time scale for the spin dynamics, e.g. the excitation transfer discussed in Sec. \ref{txt:dressed_ion_interaction}, thus increases proportional to $|c_\alpha|^{-4}$. Hence, comparing the scaling of the interaction strength and the lifetime implies that the adiabatic admixture of Rydberg states must not be too weak in order to avoid that decoherence due to spontaneous emission becomes an issue during the temporal evolution of the spin dynamics.
\subsection{Fast Two-Qubit Quantum Gates with Rydberg Ions}
\label{txt:fast_quantum_gates}
In most two-qubit gate schemes, which have been suggested and implemented so far with trapped ions (see.~Refs.~\cite{Cirac95,Schmidt-Kaler03,Monz08} for a few examples), the motional sidebands of the ions have to be spectroscopically resolved, which limits the achievable gate operation times, typically to the order of the inverse external trapping frequency. In order to overcome this limitation, it has been proposed to use specifically shaped trains of off-resonant laser pulses to implement geometric two-qubit ion gates with much faster gate times \cite{Ripoll03}. In the context of quantum information processing with neutral atoms, potentially fast two-qubit gates can be achieved for pairs of Rydberg atoms in an optical lattice, based on the strong and long-range dipolar interactions among the atoms \cite{Lukin01,Jaksch00,Moller08}. 

For the system of trapped Rydberg ions we have shown that despite the absence of permanent dipole moments MW-dressing fields can be used to generate strong dipole-dipole interactions between Rydberg ions. We now aim to exploit the electronic interaction Hamiltonian (\ref{eq:spin_chain_Hamiltonian}) for the implementation of a fast conditional two-qubit phase gate along the lines of the proposals developed for neutral Rydberg atoms. Such gate is characterized by the truth table $|g_a \rangle_m |g_b \rangle_n \rightarrow e^{i (a-2)(b-2)\phi_\mathrm{ent}} |g_a \rangle_m | g_b\rangle_n$ with $a,b=1,2$ labeling the two ground states, and the ion indices $m,n$. Thus, the two-qubit state $|g_1\rangle_m |g_1\rangle_n$ obtains a phase shift, while the other three states are unaffected (up to trivial single qubit phases) \cite{NielsenChuang}.

\begin{figure}\center
\includegraphics[angle=0,width=8cm]{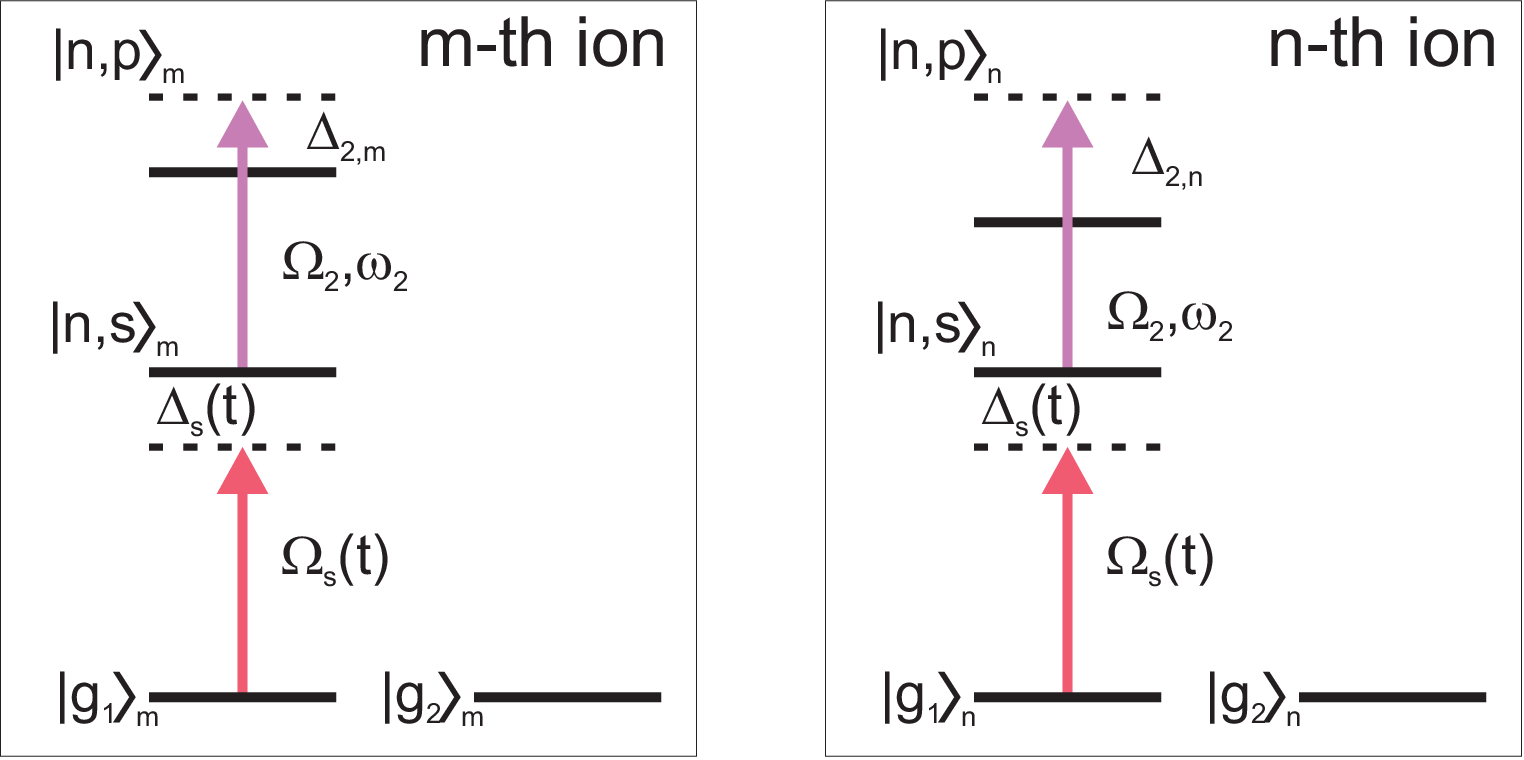}
\caption{Level scheme for the implementation of a conditional two-qubit phase gate. The ground states $|g_1\rangle$ of the respective ions are coupled via a laser with time-dependent detuning $\Delta_s(t)$ and Rabi frequency $\Omega_s(t)$ to the Rydberg level $|n,s\rangle$. A MW field of constant Rabi frequency $\Omega_2$ and ion-dependent detuning $\Delta_{2,m}$ couples the Rydberg levels $|n,s\rangle$ and $|n,p\rangle$.
}
\label{fig:gate}
\end{figure}

\begin{figure}\center
\includegraphics[angle=0,width=6cm]{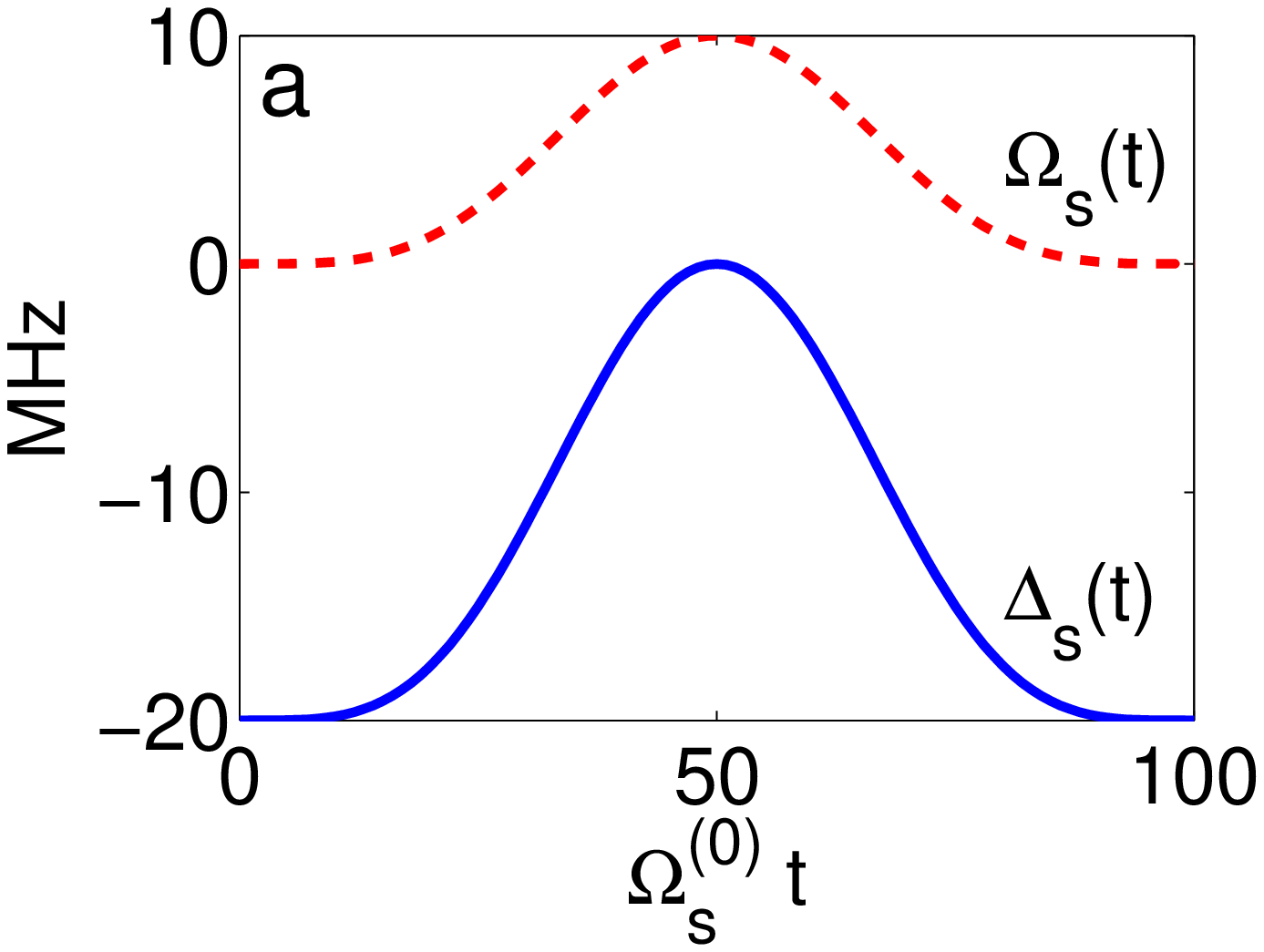}\includegraphics[angle=0,width=6cm]{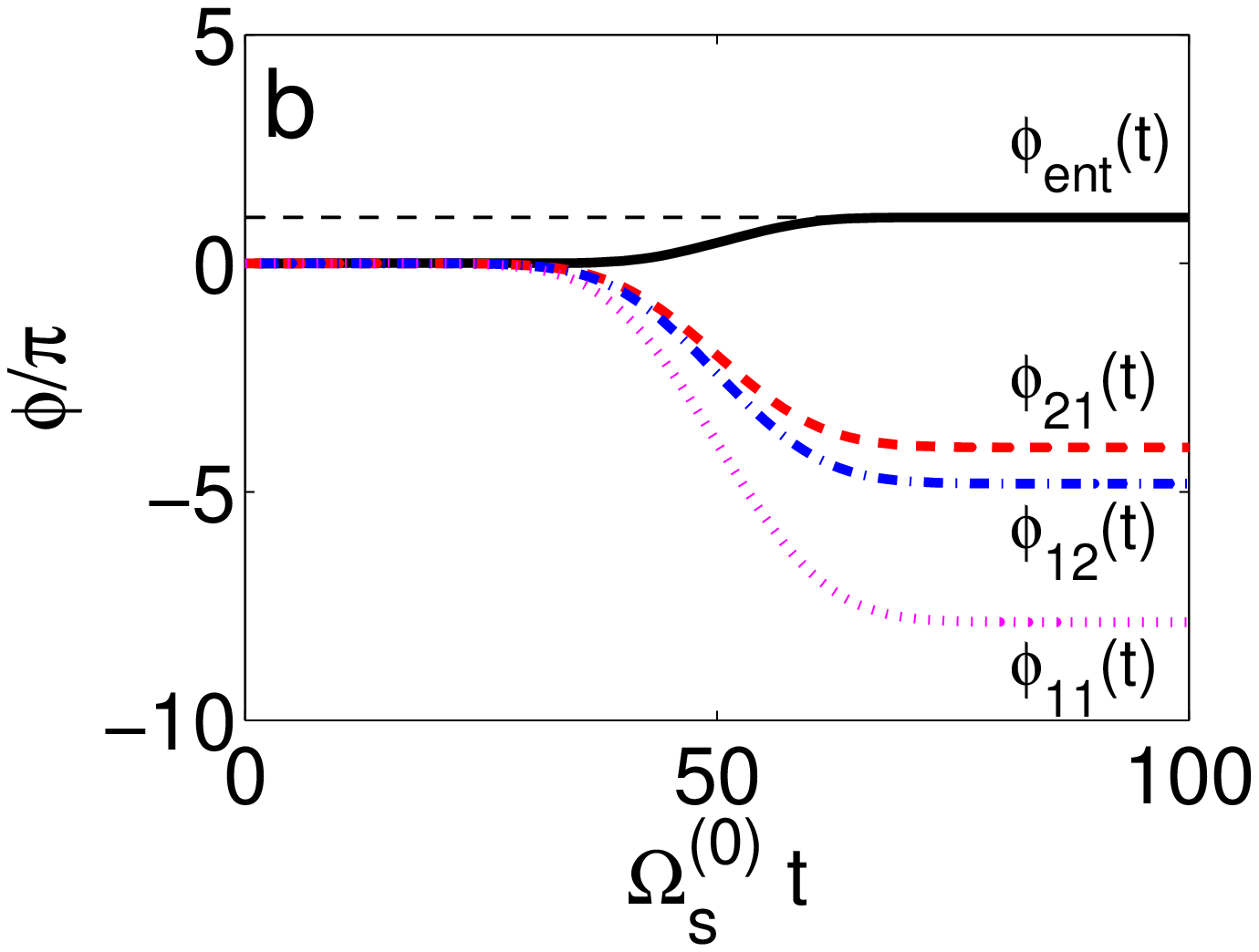}
\caption{Two-qubit conditional phase gate between the first and second ion in a chain of ten ions. \textbf{a}: Pulse profile: Rabi frequency $\Omega_s(t)$ and detuning $\Delta_s(t)$ of the dressing lasers. The MW Rabi frequency is chosen $\Omega_2=57.5$ MHz, the ion-dependent MW detunings are $\Delta_{2,1}=-279$ MHz, $\Delta_{2,2}=-667$ MHz. Dipole-dipole interaction energy scale $J/\hbar=500$ MHz (cf. Sec. \ref{txt:dressed_ion_interaction}). \textbf{b}: Accumulated phase shifts in the dressed states adiabatically connected to the initial ground states, and resulting entanglement phase $\phi_\mathrm{ent}(t)$.
}
\label{fig:gate_detuning_phase}
\end{figure}

We identify the two ground states $|g_1\rangle_m$ and $|g_2\rangle_m$ of each ion $m$ as logical qubit states $|0\rangle$ and $|1\rangle$. The ground state $|g_1\rangle$ is coupled to the Rydberg state $|n,s\rangle$ by a near-resonant laser with time-dependent effective Rabi frequency $\Omega_s(t)$ and detuning $\Delta_s(t)= \omega_s(t) -(E_{|n,s\rangle}-E_{|g_1\rangle})/\hbar$, which can be chosen equal for both ions $m$ and $n$. The second ground state $|g_2\rangle$ is not coupled to any Rydberg state. We again consider the scenario in which only one additional MW field with Rabi frequency $\Omega_2$ and ion-dependent detunings $\Delta_{2,m}$ (cf.~Sec.~\ref{txt:dressed_ion_interaction}) is applied (see Fig.~\ref{fig:gate}), i.e.~$\Omega_1=0$ and $\eta_m=0$ in Eq.~(\ref{eq:spin_chain_Hamiltonian}).
To perform the gate operation, we apply laser pulses to the two ions (similarly to the adiabatic gate scheme presented in Ref.~\cite{Jaksch00}). The variation of the laser pulses is assumed to be slow on the time-scale set by $\Omega_s$ and $\Delta_s$ such that the system follows adiabatically the dressed states, which arise from slowly switching on the laser coupling. This adiabaticity condition guarantees that after applying the laser pulses the electronic population still completely resides in the initial electronic ground states. During the application of the laser pulses and the resulting dressing of the electronic ground states part of the electronic population is transferred from the states $|g_1\rangle_m$ and $|g_1\rangle_n$ to the Rydberg states, where the ions interact via the resonant dipole-dipole interaction and thereby accumulate an entanglement phase, given by
\begin{equation}
\phi_\mathrm{ent}(t) = \int_0^t \mathrm{d}\tau (\epsilon_{|g_1\rangle|g_1\rangle}(\tau) - \epsilon_{|g_1\rangle|g_2\rangle}(\tau) - \epsilon_{|g_2\rangle|g_1\rangle}(\tau))
\end{equation}
where $\epsilon_{|g_a\rangle_m |g_b\rangle_n}$ denotes the eigenenergy of the instantaneous eigenstate connected to the state $|g_a\rangle_m |g_b\rangle_n$ in the absence of laser pulses. Fig.~\ref{fig:gate_detuning_phase}a shows a specific choice for the pulse profile, the time-dependent energies of the dressed states adiabatically connected to the different ground states. The resulting accumulated phase shifts are presented in Fig.~\ref{fig:gate_detuning_phase}b.
The gate operation time $T$ is approximately two orders of magnitude larger than the inverse of the maximum Rabi frequency $\Omega_s^{(0)}$ of the applied lasers in order to satisfy the adiabaticity condition (for the chosen set of parameters $T=100 / \Omega_s^{(0)}=10 \mu$s). In order to minimize imperfections in the gate operation due to spontaneous scattering of photons of the dressing lasers (cf.~discussion in Sec.~\ref{txt:adiabatic_scheme}), the gate operation has to take place on a time-scale much faster than the  lifetime $\tau$ of the involved Rydberg levels, i.e. $T \ll \tau$. For sufficiently high laser Rabi frequencies and long-lived Rydberg levels this condition can be satisfied.
\section{Conclusions and Outlook}
In this paper we have shown that trapping of Rydberg ions in a linear electric ion trap under realistic conditions is feasible. We have found that for not too large principal quantum numbers field ionization of the Rydberg ions due to the trapping fields is negligible, and that the coupling of electronic and external dynamics of the ions results in renormalized trapping frequencies for ions excited to Rydberg states. We have suggested to use MW dressing fields in order to generate strong dipolar interactions among the ions. The Rydberg excitation dynamics of the MW-dressed ions can be described by an effective interacting spin-1/2 Hamiltonian. The strong interactions give rise to a typical corresponding time scale of the order a few ns, which is substantially shorter than the typical decoherence time set by the radiative decay of Rydberg states. We have studied the dynamical transfer of a Rydberg excitation through the ion chain and discussed the implementation of fast two-qubit gates. While the laser excitation of ions to Rydberg states is an experimentally challenging task, the system offers the prospect of studying coherent many-body quantum dynamics on fast time scales in a well-controlled and structured environment. 

Beyond the present work, trapped Rydberg ions offer a rich playground for further studies of more involved Rydberg dynamics: Combining e.g. ions of different species or exciting only a certain number of the trapped ions to Rydberg states offers the possibility to introduce further inhomogeneities in the spatial distribution of effective spins and to study in a well-controlled way spin chain dynamics in the presence of disorder and effective spin vacancies. The analysis can also be readily extended to study Rydberg excitation dynamics in two- or three-dimensional ion crystals. The invidual addressability of  trapped Rydberg ions allows for the setup of an excitation trap \cite{Mulken07}, where an effective spin excitation propagating along the ion chain is trapped or extracted from the system once it reaches a dedicated site.

The present work focuses on the parameter regime, where the interparticle distance in the ion chain is much larger than the extension of the electronic wave function of ions excited to Rydberg states. Another interesting scenario is the situation where those two length scales become comparable. If this regime can be achieved, electrons could hop between overlapping Rydberg orbitals of adjacent ions, which would pave the way towards the realization of mesoscopic Hubbard-type models in an ion trap.

We thank Hartmut H\"affner, Piet Schmidt and Rainer Blatt for
discussions, and in particular Jonathan Home for valuable comments. This work was supported by the Austrian Science Foundation, the EU under grants CONQUEST, MRTN-CT-2003-505089, SCALA IST-15714, and the Institute for Quantum Information. L.-M. Liang
acknowledges funding by National Funds of Natural Science (Grant
No.10504042).

\appendix

\section{Hamiltonian of N Interacting Rydberg Ions}\label{app:interacting_ions}
In this appendix we derive the Hamiltonian for a system of $N$ interacting trapped Rydberg ions. Using the single ion Hamiltonian (\ref{eq:SingleIon}) and introducing a label for the respective ion we find
\begin{eqnarray}
  H_N=\sum_i^N H^\prime_i+\frac{1}{2}\sum_{i,j(\neq i)}^N V(\mathbf{R}_i,\mathbf{R}_j,\mathbf{r}_i,\mathbf{r}_j).
\end{eqnarray}
with $V$ given by Eq.~(\ref{eq:multipole_expansion}). The external dynamics is governed by the interplay of the harmonic confinement and the Coulomb force between the ions. The corresponding potential reads
\begin{eqnarray}
  V_\mathrm{ext}=\frac{1}{2} M \sum_{i}^N \left[ \omega_z^2 Z_i^2 + \omega_\rho^2 (X_i^2 +Y_i^2) \right]+\frac{1}{2} \frac{e^2}{4 \pi \epsilon_0} \sum_{i,j(\neq i)}^N\frac{1}{R_{ij}}.
\end{eqnarray}
Following Refs.~\cite{Porras04,Deng05} we perform a harmonic expansion of $V_\mathrm{ext}$ which puts us into position to describe the external dynamics in terms of uncoupled harmonic oscillators/phonon modes. To this end we first calculate the ionic equilibrium positions $\mathbf{R}^{(0)}_i$. In a linear Paul trap with tight transversal confining, $\omega_\rho \gg \omega_z$, the ions align along the $z$-axis of the trap such that $\mathbf{R}^{(0)}_i=(0,0,Z^{(0)}_i)$ where the $Z^{(0)}_i$ are determined by $(\partial V_\mathrm{ext}/\partial Z_i)=0$ which leads to the system of equations
\begin{eqnarray}
  M \omega_z^2 Z_i^{(0)} & = & \frac{e^2}{4 \pi \epsilon_0} \sum_{j(\neq i)}^N \frac{Z_i^{(0)}-Z_j^{(0)}}{|Z_i^{(0)}-Z_j^{(0)}|^3}.\label{eq:equ_condition}
\end{eqnarray}
With the $Z^{(0)}_i$ being determined, the harmonic approximation of the potential $V_\mathrm{ext}$ reads
\begin{eqnarray}
  V_\mathrm{ext} & = & \frac{1}{2} M \sum_{\alpha={x,y,z}} \sum_{i,j}^N K_{ij}^\alpha q_i^\alpha q_j^\alpha
\end{eqnarray}
with the displacements $q_i^{x}=X_i$, $q_i^{y}=Y_i$ and $q_i^{z}=Z_i-Z_i^{(0)}$ and the coefficient matrix
\begin{eqnarray}
  K_{ij}^\alpha & = & \left\{ \begin{array} {rcl} \omega_\alpha^2 -c_\alpha \frac{e^2}{4 \pi \epsilon_0 M} \sum_{k (\neq i)}^N \frac{1}{| Z_i^{(0)} - Z_k^{(0)}|^3} & \mathrm{for} & i = j \nonumber \\
c_\alpha \frac{e^2}{4 \pi \epsilon_0 M} \frac{1}{|Z_i^{(0)}-Z_j^{(0)}|^3} & \mathrm{for} & i \neq j
\end{array}\right.
\label{eq:coefficient_matrix}
\end{eqnarray}
where $c_{x,y}=1, \,\,c_z=-2$. The vibrational dynamics of the 1D-chain is then described by
\begin{equation}
H_\mathrm{ph} = \frac{1}{2} M \sum_{\alpha=x,y,z} \sum_{i,j}^N K_{ij}^\alpha q_i^\alpha q_j^\alpha + \frac{1}{2M} \sum_{\alpha=x,y,z}\sum_{i}^N (P_i^\alpha)^2.
\label{txt:H_ph_1_appendix}
\end{equation}
The Hamiltonian $H_\mathrm{ph}$ can be brought into diagonal form by introducing phonon modes via the orthogonal matrices $M^\alpha$,
\begin{eqnarray}
q_i^\alpha & = & \sum_n^N \frac{M_{i,n}^\alpha}{\sqrt{2 M \omega_{\alpha,n}/\hbar}} (a_{\alpha,n}^\dagger + a_{\alpha,n}), \\
P_k^\alpha & = & i \sum_n^N \frac{M_{k,n}^\alpha}{\sqrt{2 /(\hbar M \omega_{\alpha,n})}} (a_{\alpha,n}^\dagger - a_{\alpha,n}), \\
& & \sum_{i,j}^N M_{i,n}^\alpha K_{ij}^\alpha M_{j,m}^\alpha = \omega_{\alpha,n}^2 \delta_{n,m},
\label{eq:M_matrices_appendix}
\end{eqnarray}
which leads to Eq.~(\ref{eq:full_N_ion_Hamiltonian}) for the phonon dynamics.

We now proceed by expanding the charge-dipole, the charge-quadrupole and the dipole-dipole interaction around the equilibrium positions of the ions. The charge-quadruple and the dipole-dipole interaction can be approximated by
\begin{eqnarray}
H_\mathrm{cq} & = & \frac{1}{2}\frac{e^2}{4 \pi \epsilon_0} \sum_{i \neq j}^N \frac{r_i^2-3 (\mathbf{n}_{ij} \cdot \mathbf{r}_i)^2}{2R_{ij}^3} \simeq \frac{1}{2}\frac{e^2}{4 \pi \epsilon_0} \sum_{i \neq j}^N \frac{r_i^2-3 z_i^2}{2 | Z_i^{(0)} - Z_j^{(0)}|^3} \\
H_\mathrm{dd} & = & \frac{1}{2}\frac{e^2}{4 \pi \epsilon_0} \sum_{i\neq j}^N \frac{\mathbf{r}_i \cdot \mathbf{r}_j - 3 (\mathbf{n}_{ij} \cdot \mathbf{r}_i) (\mathbf{n}_{ij} \cdot \mathbf{r}_j) }{R_{ij}^3} \simeq \frac{1}{2}\frac{e^2}{4 \pi \epsilon_0} \sum_{i\neq j}^N \frac{\mathbf{r}_i \cdot \mathbf{r}_j - 3 z_i z_j}{| Z_i^{(0)} - Z_j^{(0)}|^3}.\nonumber \\
\end{eqnarray}
The charge-quadrupole term leads a position dependent variation of the electric field and can be absorbed in the single particle ion-field interaction $H_\mathrm{ef}$ which is given by Eq.~(\ref{eq:ion-field interaction}). The electronic Hamiltonian of the $i$-th ion then assumes the form of Eq.~(\ref{eq:internal_Ham_including_quad}) with the ion-dependent gradient (\ref{eq:position_dependent_gradient}).

In order to treat the charge-dipole coupling we introduce the compact notation $r_i^x=x_i, R_i^x=X_i, \ldots$ which enables us to write
\begin{eqnarray}
H_\mathrm{cd} & = & \frac{1}{2}\frac{e^2}{4 \pi \epsilon_0} \sum_{i\neq j}^N \frac{(\mathbf{R}_i-\mathbf{R}_j) (\mathbf{r}_i -\mathbf{r}_j)}{R_{ij}^3}  \\
& \simeq & \frac{1}{2}\frac{e^2}{4 \pi \epsilon_0} \sum_{i\neq j }^N \frac{Z_i^{(0)}-Z_j^{(0)}}{|Z_i^{(0)}-Z_j^{(0)}|^3} (z_i-z_j)\nonumber\\
&&+\sum_{\alpha=x,y,z}  \sum_{i,j(\neq i)}^N \frac{e^2}{4 \pi \epsilon_0} \frac{1}{|Z_i^{(0)}-Z_j^{(0)}|^3} c_\alpha (r^\alpha_i -r^\alpha_j) q_i^{\alpha}\nonumber.
\end{eqnarray}
It turns out that by making use of the equilibrium condition (\ref{eq:equ_condition}) one can combine this expression with the term $\sum_i H_{\mathrm{CM-el},i}$ that arises from the summation of the single ion terms of the Hamiltonian (\ref{eq:SingleIon}). The combination then includes all couplings between the internal and external dynamics. It reads
\begin{equation}
H_{\mathrm{int-ext}} = - 2 e \alpha \cos(\omega t) \sum_i^N [q_i^x x_i - q_i^y y_i] +\frac{1}{2} M \sum_{\alpha=x,y,z} \sum_{i,j}^N K_{ij}^z c_\alpha r_j^\alpha q_i^\alpha.\label{eq:int_ext_coupling}
\end{equation}
Finally, neglecting the micro-motion, the complete Hamiltonian of $N$ Rydberg ions in the linear Paul trap is given by Eq. (\ref{eq:full_N_ion_Hamiltonian}).

\section{Effective Dipole Moment of Laser-Dressed Ground State Ions}
\label{txt:dipole_moment_ground_states}
In this appendix we derive the form (\ref{eq:dipole_operator_ground_states}) of the dipole operator for laser-dressed ground state ions. The Hamiltonian for the system of five coupled electronic levels as depicted in Fig.~(\ref{fig:MW_dressing_ground_states}) in the rotating frame defined by the transformation
\begin{eqnarray}
U(t) & = &
e^{-i\omega_s t} |g_1\rangle \langle g_1| + e^{i(\omega_{2}-\omega_p)t}  |g_2\rangle \langle g_2| \nonumber \\
& & + e^{-i\omega_1 t}  |n',p \rangle \langle n',p| +  |n,s \rangle \langle n,s| + e^{i \omega_2 t}  |n,p\rangle \langle n,p|
\end{eqnarray}
and in rotating-wave-approximation is given by $H_\mathrm{5levels} = H_0 + H_\mathrm{pert}$,
\begin{eqnarray}
H_0 & = & \hbar \Delta_s  |g_1\rangle \langle g_1| + \hbar \Delta_{p}  |g_2\rangle \langle g_2| + \hbar \Delta_{1}  |n',p\rangle \langle n',p| - \hbar \Delta_2 |n,p\rangle \langle n,p| \nonumber \\
&& + \frac{\hbar}{2} \left( \Omega_1  |n',p\rangle \langle n,s | + \Omega_2  |n,s\rangle \langle n,p| + \mathrm{ h.c.} \right), \\
H_\mathrm{pert}  & = & \frac{\hbar}{2} \left( \Omega_s  |g_1\rangle \langle n,s| +\Omega_p  |g_2\rangle \langle n,p| + \mathrm{ h.c.}  \right).
\end{eqnarray}
The MW detunings $\Delta_{1,2}$ and Rabi frequencies $\Omega_{1,2}$ are defined as in Sec.~\ref{txt:dressed_rydberg_ions}, and the detunings and Rabi frequencies characterizing the two additional laser fields are given by $\Delta_s=\omega_s -(E_{|n,s\rangle}-E_{|g_1\rangle})/\hbar$, $\Delta_p=\omega_p -(E_{|n,p\rangle}-E_{|g_2\rangle})/\hbar-\Delta_2$ and $\Omega_s$, $\Omega_p$. We assume that the near-resonant MW fields are much stronger than the two lasers such that they determine the level structure of the dressed Rydberg states. Therefore we treat the coupling to the ground states as a perturbation $H_\mathrm{pert}$ of the system. We now apply a canonical transformation to the Hamiltonian \cite{Czycholl},
\begin{equation}
e^{-S} H_\mathrm{5levels}\, e^{S} =  H_0 + H_\mathrm{pert} + [H_0,S] + [H_\mathrm{pert},S] +\ldots
\end{equation}
Choosing $S$ such that $[H_0,S]=-H_\mathrm{pert}$ guarantees that to first order in the perturbation $H_\mathrm{pert}$ the transformed Hamiltonian is block-diagonal and that the two ground states become decoupled from the Rydberg states.
The transformation yields two dressed ground states, $|g_{1,2}'\rangle \simeq (1+ S)|g_{1,2}\rangle$,
\begin{eqnarray}
|g_1' \rangle & = & |g_1\rangle + c_{p',1} |n',p \rangle + c_{s,1} |n,s\rangle + c_{p,1} |n,p\rangle \\
|g_2' \rangle & = & |g_2\rangle + c_{p',2} |n',p\rangle + c_{s,2} |n,s\rangle + c_{p,2} |n,p\rangle
\end{eqnarray}
with
\begin{eqnarray}
\left( \begin{array}{c}
c_{p',1} \\ c_{s,1} \\ c_{p,1}
\end{array}\right) & = &
\gamma_s\left( \begin{array}{c}
-(\Delta_s+\Delta_2)\Omega_1 \\
2 (\Delta_1-\Delta_s)(\Delta_s+\Delta_2) \\
(\Delta_1-\Delta_s) \Omega_2
\end{array}\right) \nonumber \\
\left( \begin{array}{c}
c_{p',2} \\ c_{s,2} \\ c_{p,2}
\end{array}\right) & = &
\gamma_p \left( \begin{array}{c}
-\Omega_1 \Omega_2\\
2(\Delta_1-\Delta_p) \Omega_2\\
4(\Delta_1-\Delta_p)\Delta_p + \Omega_1^2
\end{array}\right)
\label{eq:admixture_coefficients}
\end{eqnarray}
where
\begin{eqnarray}
  \gamma_{s,p}=
\frac{\Omega_{s,p}}{(\Delta_{s,p}+\Delta_2)[4(\Delta_1-\Delta_{s,p}) \Delta_{s,p} +\Omega_1^2] + (\Delta_{s,p}-\Delta_1)\Omega_2^2}.
\end{eqnarray}
For the special choice of laser detunings
\begin{equation}
\Delta_s = - \Delta_2, \qquad \Delta_p = \frac{1}{2} \left( \Delta_1 -\sqrt{\Delta_1^2+\Omega_1^2}\right)
\end{equation}
we find
\begin{equation}
\left( \begin{array}{c}
c_{p',1} \\ c_{s,1} \\ c_{p,1}
\end{array}\right)  =
\left( \begin{array}{c} 0 \\ 0 \\ -\frac{\Omega_s}{\Omega_2}
\end{array}\right), \qquad
\left( \begin{array}{c}
c_{p',2} \\ c_{s,2} \\ c_{p,2}
\end{array}\right) =
\left( \begin{array}{c}
\frac{\Omega_1\Omega_p}{\Omega_2\left( \Delta_1 +\sqrt{\Delta_1^2+\Omega_1^2}\right)} \\
-\frac{\Omega_p}{\Omega_2} \\
0
\end{array}\right). \nonumber \\
\end{equation}
Thus, for this set of parameters, the Rydberg state $|n,p\rangle$ is exclusively admixed to the ground state $|g_1\rangle$, while the second ground state obtains a small fraction of the Rydberg states $|n',p\rangle$ and $|n,s\rangle$. This implies that the dressed ground state $|g_2'\rangle$ possesses a non-vanishing dipole moment, which oscillates at the MW frequency $\omega_1$, and that there exists also a transition dipole matrix element between the two dressed ground states. We now identify the two dressed ground states $|g_1'\rangle$, $|g_2'\rangle$ with the eigenstates of the $S_z$ spin operator along the lines of Sec.~\ref{txt:dressed_rydberg_ions} and obtain the dipole operator (\ref{eq:dipole_operator_ground_states}).
\\

\end{document}